\documentclass[11pt]{article}
\usepackage{pstricks,pst-node,pst-plot}
\usepackage{mathrsfs,amsmath,amsbsy,amssymb, mcite}

\newcommand{\e}{\epsilon}
\renewcommand{\t}{\eta}
\newcommand{\beq}{\begin{equation}}
\newcommand{\eeq}{\end{equation}}
\newcommand{\beqn}{\begin{equation*}}
\newcommand{\eeqn}{\end{equation*}}

\newcommand{\p}{\partial}
\newcommand{\w}{\wedge}

\title{Spinor classification of the Weyl tensor in five dimensions}

\author{Mahdi Godazgar\\Department of Applied Mathematics and Theoretical Physics \\ Centre for Mathematical Sciences \\ Wilberforce Road, Cambridge CB3 0WA, UK\\ mmg31@cam.ac.uk}

\begin{document}
\maketitle
\begin{abstract}
We investigate the spinor classification of the Weyl tensor in five dimensions due to De Smet. We show that a previously overlooked reality condition reduces the number of possible types in the classification. We classify all vacuum solutions belonging to the most special algebraic type. The connection between this spinor and the tensor classification due to Coley, Milson, Pravda and Pravdov{\'a} is investigated and the relation between most of the types in each of the classifications is given. We show that the black ring is algebraically general in the spinor classification.
\end{abstract}

\section{Introduction}

Despite growing interest in the study of higher dimensional gravity in recent years, most higher dimensional solutions found to date have been direct generalisations of four dimensional solutions. One way to investigate higher dimensional gravity independent of the 4d case is to attempt a systematic study of $d>4$ general relativity.  In four dimensions, the Petrov classification \cite{cartan,*ruse,*pet,*debever,*pen1,*pirani, steph} of the Weyl tensor has been a useful tool in studying solutions to the Einstein equations (a well-known example being \cite{kinn}).  Thus, it is natural to consider higher dimensional generalisations of the Petrov classification. 

A tensorial classification scheme based on a higher dimensional generalisation of the concept of principal null directions of the Petrov classification has been proposed by Coley, Milson, Pravda and Pravdov{\'a} \cite{collet, milson} (henceforth abbreviated to CMPP). The CMPP classification applies to spacetimes of any dimension.

The CMPP classification scheme has been successfully applied to studying many aspects of higher dimensional gravity (see \cite{coley} and references therein), including a partial generalisation of the Goldberg-Sachs theorem to higher dimensions \cite{durkee} (see also \cite{pravbian}), the asymptotic properties of higher dimensional spacetimes \cite{highpeel,highpeel2} and a classification of axisymmetric solutions to vacuum Einstein equations in higher dimensions \cite{axi}.

Another higher dimensional classification scheme has been proposed by De Smet \cite{smet}.  The De Smet classification generalises the concept of a Petrov Weyl spinor to five dimensions.  The classification uses the 5d Clifford algebra to define a totally symmetric 4-spinor, called the \textit{Weyl spinor}, that is equivalent to the Weyl tensor.  A given solution is classified by studying how its Weyl spinor factorises.  Or more precisely, how the fourth order quartic homogeneous polynomial formed from its Weyl spinor factorises.  The fact that the Weyl spinor is generally complex means that it must satisfy a reality condition and this reduces the number of possible types.

In five dimensions, it is known that the two classification schemes are not equivalent; that is they do not agree on the definition of an ``algebraically special'' solution. An example is known that is algebraically special in the CMPP classification, but algebraically general in the De Smet classification \cite{her} and vice versa \cite{axi}.  The presence of two inequivalent classification schemes in five dimensions presents us with the opportunity of studying solutions that are algebraically general in one scheme and special in the other.

Apart from a classification of static axisymmetric solutions belonging to two particular algebraic types \cite{smet,desmet22}, the De Smet classification has not been studied much.  The aim of this paper is to better understand the De Smet classification and its relation to the 4d Petrov and 5d CMPP classifications.  We shall find that the previously overlooked reality condition will play an important part in this study.

As a way of highlighting the most important characteristics of the spinor classification of the Weyl tensor, we shall also consider the spinor classification of two-form fields, where it is much easier to appreciate subtle issues such as reality conditions.  This is because, we shall be dealing only with a bispinor, rather than a 4-spinor as is the case in the Weyl classification.

Therefore, we begin, in section \ref{2form}, with a derivation of a spinor classification of 2-form fields.  We construct a bispinor equivalent of the 2-form and use properties of the Clifford algebra to show that it is symmetric and satisfies a reality condition.  This leads to a classification of 2-forms based on whether the equivalent bispinor factorises or not.

Then, in section \ref{sec:smet}, we move on to derive the spinor classification of the Weyl tensor due to De Smet \cite{smet} in similar vein to the derivation of the spinor classification of 2-forms in section \ref{2form}. We define the Weyl spinor, and show that it is totally symmetric and satisfies a reality condition. The reality condition reduces the number of algebraically special types.

The De Smet classification is intended to be a generalisation of the spinor formulation of the Petrov classification to five dimensions.  It is not clear, though, how the two schemes are related and in what sense the De Smet classification is a generalisation of Petrov's beyond the superficial link that they both deal with the factorisability properties of a totally symmetric spinor quantity.  This issue is addressed in section \ref{smet4d}, where it is shown that one can define an analogue of the De Smet Weyl spinor in 4d and that the classification of the Weyl tensor based on this can be thought of as a classification using Majorana spinors.  Recall that in the Petrov classification, one uses chiral spinors.

In section \ref{prod}, we use the results obtained in section \ref{smet4d} to study direct product solutions.  We find that the De Smet type of solutions with a 4d factor is equal to the De Smet type of the 4d submanifold.  Thus, the analysis reduces to that done in section \ref{smet4d}.  For the case with 2d and 3d Lorentzian factors with non-zero cosmological constant, the Weyl spinor factorises into two proportional bispinors that cannot be further factorised.  These results are similar to those found in the study of warped product manifolds in the context of the CMPP classification in \cite{typed}.

In section \ref{2formcomp}, we consider the connection between the tensor and spinor classifications of a 2-form, where the tensor classification is based on the CMPP classification.  We find that a solution of any spinor algebraic type may be algebraically general in the tensorial sense.  For solutions that are algebraically special in the spinorial sense, what determines whether they are algebraically special or general in the tensorial sense is whether the vector that can be formed from the spinor that we have from the factorisation of the bispinor is null or timelike.  

We find that similar statements can be made regarding the relation between the De Smet and 5d CMPP classifications of the Weyl tensor in section \ref{CMPPSmet}.  However, in this case we cannot study all types fully.  Thus, we begin by assuming that the solution is of type N, III or D and derive the general De Smet polynomials for the respective cases.  Considering the factorisability properties of these general polynomials gives the possible spinor types that they can have.  We show that type III and D solutions may be algebraically general in the spinor classification, while for type N solutions, the De Smet polynomial is guaranteed to factorise into linear factors, so type N solutions are also algebraically special in the spinor classification.  We do not consider more general types due to the complexities of factorising a general polynomial. Then, we go on to consider the reverse case, i.e. assuming a particular De Smet type and examining what this implies about the CMPP type. Since the general form of Weyl tensor is important for this analysis, we can only do this for the case where the Weyl spinor factorises into two bispinors, or a more special case of this, using the general form of the Weyl tensor of such solutions derived in section \ref{sec:smet}.  Thus, we do not consider the case where the solution is algebraically general, i.e. the Weyl spinor does not factorise nor the case where it factorises into a rank-3 spinor and a univalent spinor.  We find that any spinor type may be algebraically general in the CMPP sense.

An important motivation, given above, for understanding the De Smet classification and its relation to the 5d CMPP classification was that this may allow us to study 5d solutions that are algebraically general in one classification scheme and special in the other. Furthermore, the result found in section \ref{CMPPSmet} that any spinor type may be algebraically general in the CMPP sense strengthens this motivation.  The black ring \cite{br} is a well-known example of a CMPP algebraically general five dimensional solution \cite{prav}\footnote{In \cite{prav}, it is shown that WANDs can only be found in certain regions for the black ring and it is claimed that the black ring is type I.  However, if we take the strict definition of the classification, which states that the algebraic type of the spacetime corresponds to the type of its most algebraically general point, then the black ring is type G. The black ring is an example of a solution that is type G in one open region and type I in another.  This kind of behaviour is discussed in \cite{axi}.}.  Therefore, it would be desirable to know the De Smet type of the black ring solution.  It is shown in section \ref{br} that the black ring is also, unfortunately, algebraically general in the De Smet classification.

In section \ref{em}, we consider the constraints imposed on a spacetime by the existence of an algebraically special 2-form solving Maxwell-type equations. An algebraically special 2-form is defined by a single spinor.  In 4d, the existence of an algebraically special Maxwell field is equivalent to the spacetime being algebraically special.  This follows from the Mariot-Robinson \cite{mariot,*robinson, steph} and Goldberg-Sachs \cite{gold} thereoms.  Thus, studying the existence of algebraically special fields can shed light on the status of the Goldberg-Sachs theorem in higher dimensions. The 2-form field is assumed to satisfy the Bianchi identity and a general equation of motion that includes Maxwell theory as well as minimal supergravity \cite{minsugra}.  The analysis splits into two cases of whether the vector derived from the spinor that defines the 2-form field is null or timelike.  From section \ref{2formcomp}, we know that if it is null then this is equivalent to the field being algebraically special in the CMPP sense.  This analysis has already been done in \cite{ghp}. The null vector defines a geodesic congruence with constraints on its optical properties, which are explained in section \ref{em}.  If the vector is timelike, then the solution admits a timelike geodesic congruence and an almost-K\"ahler structure.

Finally, in section \ref{1spin}, we undertake a classification of solutions belonging to the most special type, that is type $\underline{11} \ \underline{11}$ solutions.  These are defined as those for which the Weyl spinor factorises into two proportional bispinors, which factorise further into spinors. The reality condition gives that the Weyl tensor is fully determined from a single spinor.  We use the Bianchi identity to find constraints on this spinor for a vacuum Einstein solution.  As in section \ref{em}, the analysis divides into two cases of whether the vector defined from the spinor is null or timelike.  If the vector is null, then we have a type N Kundt solution satisfying further conditions that are explained in section \ref{1spin}.  The timelike case reveals more structure.  The spacetime is found to be a cosmological solution with spatial geometry a type (D,O) Einstein solution.  The solutions in section \ref{1spin} are more constrained that those found in section \ref{em}.

The index conventions in this paper are as follows:  indices $a,b,c \dots$ refer to orthonormal or null frame basis vectors and generally take values from 0 to 4, although this is not always the case.  Indices $i,j,k \dots$ refer to spacelike basis vectors and generally take values from 2 to 4. In section \ref{CMPPtoSmet}, where we move between orthonormal and null frame bases, indices $a,b,c \dots$ refer to null frame basis vectors, while $\mu, \nu, \rho \dots$ refer to orthonormal basis vectors. $\alpha, \beta, \gamma \dots$ and $\dot{\alpha}, \dot{\beta}, \dot{\gamma} \dots$ label left and right-handed chiral spinor indices in four dimensions and run from 1 to 2, while $A, B, C \dots$ and $\dot{A}, \dot{B}, \dot{C} \dots$ label Dirac and Dirac complex conjugate indices in four and five dimensions and run from 1 to 4.  There are additional index conventions in section \ref{prod}, which are explained separately in that section.

\subsection{CMPP classification} \label{CMPP}

The CMPP classification relies on the existence of a null frame in which certain components of the Weyl tensor vanish, implying the solution to be a certain type. 

Given a null frame $(\ell,n,m_i)$, one can apply four sets of continuous Lorentz transformations: null rotations about $\ell$ and $n$, spins (rotating the spacelike basis vectors), and boosts, given by
\beqn
\ell'=\lambda\ell, \qquad n'=\lambda^{-1} n, \qquad m'_i=m_i,
\eeqn 
where $\lambda \neq0$.  Under a boost, a particular component of a $p$-rank tensor $T$ in the null frame transforms as
\beqn
T_{a_1 \dots a_p} \longrightarrow \lambda^{b} T_{a_1 \dots a_p},
\eeqn
where $b$ is the \emph{boost weight} of $T_{a_1 \dots a_p}$, and is equal to the number of $a_i$ that are 0 minus the number that are 1.

For the Weyl tensor, the possible boost weights lie in the range $-2 \leq b \leq 2$.  For example, boost weight +2 components of the Weyl tensor are $C_{0i0j}$.

The solution is type G at a point $p$ if there does not exist a $\ell$ such that $C_{0 i 0 j}=0$ at $p$, i.e. $C_{0 i 0 j}\neq0$ at $p$ for any choice of $\ell$. If there does exist a $\ell$ such that $C_{0 i 0 j}=0$ at $p$, then $\ell$ called a \emph{Weyl aligned null direction} or \emph{WAND} and the solution is type I or more special at $p$. The solution is said to be \emph{algebraically special} at $p$ and of given type
\begin{itemize}
    \item II $\iff$ $C_{0 i 0 j}=C_{0 i j k}=0,$

    \item D $\iff$ $C_{0 i 0 j}=C_{0 i j k}=C_{1 i j k}=C_{1 i 1 j}=0,$

    \item III $\iff$ $C_{0 i 0 j}=C_{0 i j k}=C_{0 1 i j}=C_{i j k l}=0$,

    \item N $\iff$ $C_{0 i 0 j}=C_{0 i j k}=C_{0 1 i j}=C_{i j k l}=C_{1 i j k}=0$,

    \item O $\iff$ $C_{abcd}=0$.
\end{itemize}

\begin{figure}[!ht]
\begin{center}
\begin{pspicture}(0,1.4)(8,3.5)
\rput(0,3){G} \rput(2,3){I} \rput(4,3){II} \rput(6,3){III} \rput(8,3){N} 
\rput(4,1){D} \rput(8,1){O}
\psline[linewidth=0.2mm]{->}(0.35,3)(1.65,3)
\psline[linewidth=0.2mm]{->}(2.35,3)(3.65,3)
\psline[linewidth=0.2mm]{->}(4.35,3)(5.65,3)
\psline[linewidth=0.2mm]{->}(6.35,3)(7.65,3)
\psline[linewidth=0.2mm]{->}(4,2.65)(4,1.35)
\psline[linewidth=0.2mm,linestyle=dashed]{->}(8,2.65)(8,1.35)
\psline[linewidth=0.2mm,linestyle=dashed]{->}(4.35,1)(7.65,1)
\end{pspicture}
\end{center}
\caption{Penrose diagram of the CMPP classification}
\label{fig:cmpp}
\end{figure}
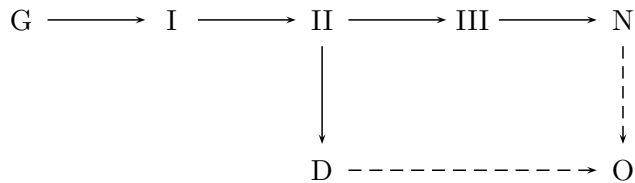

The algebraic type of the solution is defined to be the type of its most algebraically general point. If the solution is algebraically special, then $\ell$ for which $C_{0 i 0 j}=C_{0 i j k}=0$ is said to be a \emph{multiple WAND} or \emph{m}WAND.  A given \textit{m}WAND need not be unique. Indeed there may be infinitely many \textit{m}WANDs in a spacetime. For example, consider $dS_{3} \times S^{d-3}$ (for $d>4$). Any null vector in $dS_{3}$ is a \textit{m}WAND \cite{axi}.

Note that the definition of type D solutions depends on $n$ being multiply Weyl aligned as well, i.e. $C_{1i1j} = C_{1 i j k} = 0$.  Thus, the type D definition requires a secondary classification, in which $n$ is chosen such that as many trailing Weyl tensor components as is possible can be set to zero, that is, of course, once a WAND $\ell$ has been found such that as many leading Weyl tensor components as is possible have been set to zero.   For example one defines type I$_i$ solutions to be those for which a $\ell$ and $n$ can be found such that $C_{0i0j}=C_{1i1j}=0$.  We shall not utilise the secondary classification scheme here, except in the definition of type D solutions.

\section{Spinor classification of two-form and Weyl tensor}

\subsection{Spinor classification of two-form} \label{2form}

Let $F_{ab}$ be a real two-form.  We can construct a bispinor $\epsilon_{AB}$ that is equivalent to the 2-form \footnote{See appendix \ref{cliff} for conventions used for the 5d Clifford algebra.}
\beq \label{bispi}
\epsilon_{AB}=\frac{i}{8} F_{ab} {\Gamma^{ab}}_{AB},
\eeq
where $\Gamma^{a b}=\Gamma^{[a}\Gamma^{b]}$.  As explained in appendix \ref{cliff}, for brevity, we omit factors of $C$ and $C^{-1}$ where it is clear from the index structure that charge conjugation matrices have been used.  Thus, ${\Gamma^{ab}}_{AB}=(C\Gamma^{ab})_{AB}=C_{AC}{\Gamma^{ab \, C}}_{B}$.

It can be shown, using the antisymmetry of the charge conjugation matrix $C$ that $\Gamma^{a b}_{\ \ AB}$ is symmetric in its spinor indices: using the definition of $C$, we find that
\beq \label{gammtransp}
\Gamma_{a b}^{t}=-C \Gamma_{a b} C^{-1},
\eeq
which implies
\beqn
(C^t\Gamma_{a b})^t=-(C\Gamma_{a b}),
\eeqn
i.e.
\beq \label{gammasym}
(C\Gamma_{a b})^t=(C\Gamma_{a b}).
\eeq
Thus, the bispinor $\epsilon_{AB}$ is symmetric.

Using properties of $gamma$-matrices, we can invert equation \eqref{bispi}
\beq
F_{ab}=i \, tr(\Gamma_{ab} \epsilon),
\eeq
where $tr(\Gamma_{ab} \epsilon) = tr(\Gamma_{ab} C^{-1} \epsilon)$ $= {{\Gamma_{ab}}^{A}}_{B}C^{BC}\epsilon_{CA}$.

Note that while the 2-form is real, the bispinor is generally complex since there is no Majorana representation of the Clifford algebra in five dimensions. A complex bispinor has 20 real independent components, whereas a real 2-form has 10 real components. Therefore, the bispinor must satisfy a reality condition, which halves its number of independent components.

Using the definitions of the Dirac and charge conjugation matrices (see appendix \ref{cliff}) one can derive the following relation between $\Gamma_{ab}^*$ and $\Gamma_{ab}$
\beq \label{gammastar}
\Gamma_{ab}^*=A\, \Gamma_{ab} \, A^{-1},
\eeq
where $A=(CB^{-1})^t$.  Now, taking the complex conjugate of equation \eqref{bispi} and using the equation above gives
\beqn
\epsilon_{\dot{A}\dot{B}}=-\frac{i}{8} F_{ab} \, {A_{\dot{A}}}^{A}{\Gamma^{ab}}_{AB}{(A^{-1})^{B}}_{\dot{B}},
\eeqn
where ${A_{\dot{A}}}^{A}={(C A C^{-1})_{\dot{A}}}^{A}=- {(A^{-1})^{A}}_{\dot{A}}$ \cite{lee}.  Then re-arranging the above equation gives
\beq \label{realitybispi}
\epsilon_{AB}=\bar{\epsilon}_{AB},
\eeq
where $\bar{\epsilon}_{AB} \equiv \epsilon_{\dot{A}\dot{B}} {A^{\dot{A}}}_{A} {A^{\dot{B}}}_{B}$.

A 2-form field is said to be \emph{algebraically special} if the bispinor factorises. If this is the case, then the reality condition, equation \eqref{realitybispi}, implies that \footnote{A sketch of the proof of this result is given in appendix \ref{reality2form}.}
\beq
\epsilon_{AB}=\epsilon_{(A}\bar{\epsilon}_{B)}.
\eeq
Then, the 2-form $F$ is of the form
\beq
F_{ab}=i \bar{\epsilon}\Gamma_{ab} \epsilon
\eeq
We can also form a real scalar and vector 
\beq \label{scavec}
f=\bar{\epsilon}\epsilon, \qquad V^a=i \bar{\epsilon}\Gamma^a \epsilon.
\eeq
The Fierz identity can be used to relate the above three quantities \cite{minsugra}
\begin{align}
 V^2&=-f^2 \label{f1} \\
 F^2&=F^{ab}F_{ab}=4f^2 \label{f2} \\
\iota_V F&=0 \label{f3} \\
{F_a}^{c}{F_c}^{b}&=-f^2{\delta_a}^{b}-V_a V^b \label{f6} \\
F \wedge F &=2f \star V \label{f5} \\
f F&=\star(V \wedge F). \label{f4}
\end{align}
Note that the above equations are not independent. In fact, equations \eqref{f1} and \eqref{f4} can be used to derive equations \eqref{f2}--\eqref{f5}.

\subsection{De Smet classification} \label{sec:smet}
In four dimensions, the Petrov classification is most simply derived by defining a totally symmetric \emph{Weyl spinor} $\Psi_{\alpha \beta \gamma \delta}$ and considering the \emph{Weyl polynomial}
\beqn
\Psi(\chi)=\Psi_{\alpha \beta \gamma \delta} \chi^\alpha \chi^\beta \chi^\gamma \chi^\delta
\eeqn
formed from the Weyl spinor, where $\chi^\alpha$ is a general chiral spinor. The fundamental theorem of algebra ensures the factorisability of the polynomial and the Petrov classification reduces to an analysis of the multiplicity of the factors. 

In similar vein, the spinor classification of the Weyl tensor in five dimensions \cite{smet} uses a spinorial approach to the classification of the Weyl tensor.  Define the Weyl spinor, associated with the Weyl tensor, to be
\begin{equation}
    C_{ABCD}=C_{abcd} {\Gamma^{ab}}_{AB} {\Gamma^{cd}}_{CD}.
    \label{weyspi5}
\end{equation}

The Weyl spinor is symmetric in its first and last pair of indices since $C\Gamma_{ab}$ is symmetric (see section \ref{2form}). Also, using the symmetries of the Weyl tensor, it is symmetric under interchange of $AB$ and $CD$.  In five dimensions, the Fierz identity can be used to show that it is totally symmetric.  The five dimensional Fierz identity is
\beq
M_{AB}N_{CD}=\dfrac{1}{4}C_{AD} (NM)_{CB}+\dfrac{1}{4} {\Gamma_{e}}_{AD}(N\Gamma^{e}M)_{CB}-\dfrac{1}{8} {\Gamma_{ef}}_{AD}(N\Gamma^{ef}M)_{CB}.
\eeq
Letting $M=\Gamma^{ab}$ and $N={\Gamma^{cd}}$, and multiplying by $C_{abcd}$ gives
\begin{align} 
C_{abcd}{\Gamma^{ab}}_{AB} {\Gamma^{cd}}_{CD}=\dfrac{1}{4}C_{abcd}C_{AD} (\Gamma^{cd}\Gamma^{ab})_{CB}+&\dfrac{1}{4} C_{abcd}{\Gamma_{e}}_{AD}(\Gamma^{cd}\Gamma^{e}\Gamma^{ab})_{CB} \notag \\ 
-&\dfrac{1}{8}C_{abcd} {\Gamma_{ef}}_{AD}(\Gamma^{cd}\Gamma^{ef}\Gamma^{ab})_{CB}.  \label{fierz5:2}
\end{align}
The trace free property of the Weyl tensor implies that $C_{abcd}\Gamma^{a}\Gamma^{b}\Gamma^{c}=C_{abcd}\Gamma^{abc}=C_{a[bcd]}\Gamma^{abc}$.  Thus, the Bianchi identity gives
\beqn
C_{abcd}\Gamma^{a}\Gamma^{b}\Gamma^{c}=C_{abcd}\Gamma^{b}\Gamma^{c}\Gamma^{d}=0.
\eeqn
Therefore, equation \eqref{fierz5:2} reduces to
\beq \label{fierz5:wspin}
C_{ABCD}=\dfrac{1}{4} C_{abcd}{\Gamma_{e}}_{AD}([\Gamma^{cd},\Gamma^{e}]\Gamma^{ab})_{CB}-\dfrac{1}{8}C_{abcd} {\Gamma_{ef}}_{AD}([\Gamma^{cd},\Gamma^{ef}]\Gamma^{ab})_{CB}.
\eeq
Using the following identities
\beq \label{comm:gamma1}
[\Gamma^{ab},\Gamma^{c}]=2(g^{bc}\Gamma^{a}-g^{ac}\Gamma^{b}),
\eeq
\beq \label{comm:gamma2}
[\Gamma^{ab},\Gamma^{cd}]=2(g^{bc}\Gamma^{ad}+g^{ad}\Gamma^{bc}-g^{ac}\Gamma^{bd}-g^{bd}\Gamma^{ac}),
\eeq
equation \eqref{fierz5:wspin} reduces to
\beqn
C_{ABCD}=C_{ADCB}.
\eeqn
Therefore, the Weyl spinor is totally symmetric \footnote{The fact that $C_{ABCD}$ as defined by equation \eqref{weyspi5} is totally symmetric depends very much on properties of the 5d Clifford algebra and the 5d Fierz identity. At least, with regard to the antisymmetry property of $C$, which is crucial in ensuring that $C\Gamma_{ab}$ is symmetric, this does not hold in $d=7, 8, 9$ mod 8 \cite{lee}.  That is, in these dimensions, a representation of the Clifford algebra for which $C$ is antisymmetric does not exist.} 
\beq
C_{ABCD}=C_{(ABCD)}.
\eeq

As with the case of the 2-form in section \ref{2form}, the Weyl tensor is real, while the Weyl spinor will in general be complex.  The complex Weyl spinor has 70 real independent components, while the 5d Weyl tensor has 35 independent components. Thus, the Weyl spinor satisfies a reality condition, which halves its number of independent components.

Taking the complex conjugate of equation \eqref{weyspi5} and using equation \eqref{gammastar} gives
\beq \label{reality}
C_{ABCD}=C_{\dot{A}\dot{B}\dot{C}\dot{D}} {A^{\dot{A}}}_{A} {A^{\dot{B}}}_{B} {A^{\dot{C}}}_{C} {A^{\dot{D}}}_{D}.
\eeq

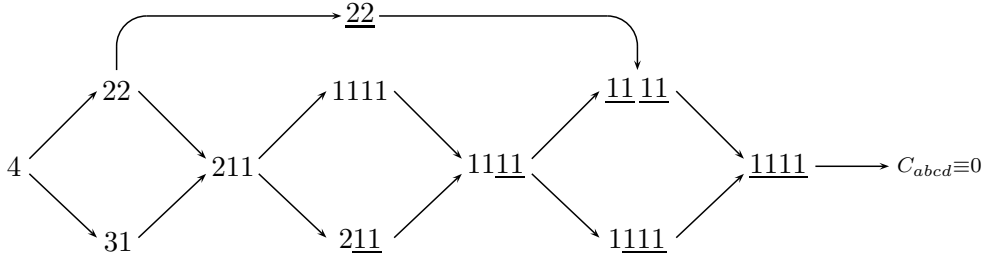
\begin{figure}[!ht]
\begin{center}
\begin{pspicture}(12.6,3.5)(0,0)
\rput(0,1){4} \rput(1.38,0){31} \rput(1.36,2){22}
\rput(2.91,1){211} \rput(4.6,0){$2\underbar{11}$}
\rput(4.6,2){1111} \rput(4.6,3){$\underbar{22}$}
\rput(6.4,1){$11\underbar{11}$} \rput(8.28,0){$1\underbar{111}$}
\rput(8.284,2){$\underbar{11} \, \underbar{11}$}
\rput(10.16,1){$\underbar{1111}$}
\rput(12.3,1){$\scriptstyle{C_{abcd}\equiv 0}$}
\psline[linewidth=0.2mm]{->}(0.2,0.9)(1.1,0.05)
\psline[linewidth=0.2mm]{->}(0.2,1.1)(1.1,2)
\psline[linewidth=0.2mm]{->}(1.65,0.05)(2.55,0.9)
\psline[linewidth=0.2mm]{->}(1.65,2)(2.55,1.1)
\psline[linewidth=0.2mm]{->}(3.25,0.9)(4.15,0.05)
\psline[linewidth=0.2mm]{->}(3.25,1.1)(4.15,2)
\psline[linewidth=0.2mm]{->}(5.05,0.05)(5.95,0.9)
\psline[linewidth=0.2mm]{->}(5.05,2)(5.95,1.1)
\psline[linewidth=0.2mm]{->}(6.88,0.9)(7.78,0.05)
\psline[linewidth=0.2mm]{->}(6.88,1.1)(7.78,2)
\psline[linewidth=0.2mm]{->}(8.8,0.05)(9.7,0.9)
\psline[linewidth=0.2mm]{->}(8.8,2)(9.7,1.1)
\psline[linewidth=0.2mm]{->}(10.65,1)(11.63,1)
\psline[linewidth=0.2mm](1.36,2.28)(1.36,2.6)
\psline[linewidth=0.2mm]{->}(1.76,3)(4.35,3)
\psline[linewidth=0.2mm](4.85,3)(7.884,3)
\psline[linewidth=0.2mm]{->}(8.284,2.6)(8.284,2.28)
\psarc[linewidth=0.2mm](1.76,2.6){0.4}{90}{180}
\psarc[linewidth=0.2mm](7.884,2.6){0.4}{0}{90}
\end{pspicture}
\end{center}
\caption{The 12 different algebraic types in the spinor classification.}
\label{fig:smet}
\end{figure}

The De Smet classification involves the factorisability properties of the invariant Weyl polynomial
\begin{equation}
    C(\psi)=C_{ABCD}\psi^A \psi^B \psi^C \psi^D,
    \label{Weylpol}
\end{equation}
where $\psi$ is a general Dirac spinor. In contrast to the Petrov classification, in general, the polynomial above will not factorise. If it does factorise, the solution is said to be \textit{algebraically special}. Each polynomial factor in the product is distinguished by its degree and multiplicity.  There are 12 possibilities, as depicted in figure \ref{fig:smet} \cite{smet}.

The notation is such made that a number represents the degree of the polynomial factor and an underline represents its multiplicity.  For example, type $\underline{22}$ corresponds to the case where the Weyl polynomial factorises into two quadratic factors that are proportional to one another and cannot be further factorised.  Type 4 solutions (for which the polynomial does not factorise) are said to be \textit{algebraically general}.

For type 22 or more special solutions, we can learn more. For all such solutions, the Weyl spinor is of the form
\beq \label{weylspi22}
C_{ABCD}=\epsilon_{(AB}\eta_{CD)}.
\eeq
The reality condition, equation \eqref{reality}, reduces to
\beq
\e_{(AB}\t_{CD)}=\bar{\e}_{(AB}\bar{\t}_{CD)},
\eeq
where
\beqn
\bar{\e}_{AB}\equiv \e_{\dot{A} \dot{B}} {A^{\dot{A}}}_{A} {A^{\dot{B}}}_{B}.
\eeqn
It can be shown that this implies that either \footnote{A sketch of the proof of this result is given in appendix \ref{reality22}.}
\beq \label{real22a}
\e_{AB} = \bar{\e}_{AB}, \qquad \t_{AB} = \bar{\t}_{AB},
\eeq
or
\beq \label{real22b}
\e_{AB} = \bar{\t}_{AB}.
\eeq

We can invert equation \eqref{weyspi5}, so that the Weyl tensor is given in terms of the Weyl spinor, i.e.
\beq \label{weytenspi5}
C_{abcd}=\frac{1}{64}(\Gamma_{ab})^{AB}(\Gamma_{cd})^{CD}C_{ABCD}.
\eeq
Using equation \eqref{weylspi22} and the 5d Fierz identity, one can derive the general form of the Weyl tensor of type 22 or more special solutions \footnote{See appendix \ref{weyl22s} for the derivation of the form of the Weyl tensor of type 22 or more special solutions.}
\begin{align} 
C_{abcd}=&A_{a[c}B_{d]b}+B_{a[c}A_{d]b} -A_{ab}B_{cd}-B_{ab}A_{cd}-\frac{1}{2} A^{ef}B_{ef}g_{a[c}g_{d]b} \notag  \\ \label{Weyl22} &-A_{ae}{B^e}_{[c}g_{d]b} -B_{ae}{A^e}_{[c}g_{d]b}+A_{be}{B^e}_{[c}g_{d]a}+B_{be}{A^e}_{[c}g_{d]a},
\end{align}
where 
\beqn
A_{ab}=i\, tr(\Gamma_{ab} \epsilon) \quad \text{and} \quad B_{ab}=i\, tr(\Gamma_{cd} \eta).
\eeqn

From the derivation of the reality condition in section \ref{2form}, we find that reality conditions \eqref{real22a} and \eqref{real22b} translate to
\beq
A^*_{ab}=A_{ab}, \quad B^*_{ab}=B_{ab}
\eeq
and
\beq
A^*_{ab}=B_{ab},
\eeq
respectively.

The reality condition constrains the Weyl spinor and we can use the results above to show that some types are not possible.

\medskip

\noindent {\bf Type $\underline{1111}$}

\medskip

The Weyl spinor of type $\underline{1111}$ solutions is of the form
\beq
C_{ABCD}=\epsilon_{A} \epsilon_{B} \epsilon_{C} \epsilon_{D}.
\eeq
Letting 
\beqn
\epsilon_{AB}=\epsilon_A \epsilon_B, \quad \eta_{AB}=\epsilon_A \epsilon_B,
\eeqn
reality conditions \eqref{real22a} and \eqref{real22b} both reduce to
\beqn
\epsilon_{A}\epsilon_{B}=\bar{\epsilon}_{A} \bar{\epsilon}_{B}.
\eeqn
In appendix \ref{reality2form}, we show that this implies a Majorana condition on $\epsilon$
\beqn
\epsilon_A \propto \bar{\epsilon}_A,
\eeqn
which gives that $\epsilon=0$ since the Majorana condition has no non-trivial solutions in 5d. $\epsilon=0$ contradicts the assumption that the solution is not conformally flat.
\medskip

\noindent {\bf Type $1\underline{111}$}

\medskip

The Weyl spinor of type $1\underbar{111}$ solutions is of the form
\beq
C_{ABCD}=\eta_{(A} \epsilon_{B} \epsilon_{C} \epsilon_{D)},
\eeq
where $\eta \not\propto \epsilon$.  We can choose
\beqn
\epsilon_{AB}=\epsilon_A \epsilon_B, \quad \eta_{AB}=\eta_{(A} \epsilon_{B)}.
\eeqn

Reality condition \eqref{real22a} gives two constraints, one of which is
\beqn
\epsilon_{A}\epsilon_{B}=\bar{\epsilon}_{A} \bar{\epsilon}_{B},
\eeqn
which contradicts the original assumption, as we showed for type $\underline{1111}$ solutions.

Reality condition \eqref{real22b} reduces to
\beq \label{1(111)_2}
\epsilon_{A}\epsilon_{B}=\bar{\epsilon}_{(A} \bar{\eta}_{B)}.
\eeq
Since $\epsilon \neq 0$, this gives
\beqn
\epsilon_{A}=\alpha \, \bar{\epsilon}_{A} + \beta \, \bar{\eta}_{A}.
\eeqn
Substituting this into equation \eqref{1(111)_2} gives
\beqn
\alpha^2 \bar{\epsilon}_{A} \bar{\epsilon}_{B} +(2 \alpha \beta -1) \bar{\epsilon}_{(A} \bar{\eta}_{B)}+\beta^2 \bar{\eta}_{A}\bar{\eta}_{B}=0.
\eeqn
Using similar techniques to those used in appendix \ref{reality2form}, it is not too difficult to show that this implies
\beqn
\alpha =0\ \quad \text{or} \quad \beta=0.
\eeqn
$\beta=0$ gives a Majorana condition on $\epsilon$, so $\alpha=0$.  Then equation \eqref{1(111)_2} becomes
\beqn
\beta^2\, \bar{\eta}_{A}\bar{\eta}_{B}=\bar{\epsilon}_{(A} \bar{\eta}_{B)}.
\eeqn
Since $\bar{\eta}\neq 0$, this implies that $\epsilon \propto \eta$.  However, this contradicts the assumption that the solution is type $1\underbar{111}$.

\medskip

\noindent {\bf Type $11\underline{11}$}

\medskip

The Weyl spinor of type $11\underbar{11}$ solutions is of the form
\beq
C_{ABCD}=\eta_{(A} \kappa_{B} \epsilon_{C} \epsilon_{D)},
\eeq
where none of the spinors are proportional to one another.  Choose
\beqn
\epsilon_{AB}=\epsilon_A \epsilon_B, \quad \eta_{AB}=\eta_{(A} \kappa_{B)}.
\eeqn

As before, reality condition \eqref{real22a} gives two constraints, one of which is
\beqn
\epsilon_{A}\epsilon_{B}=\bar{\epsilon}_{A} \bar{\epsilon}_{B},
\eeqn
which gives a contradiction.

Reality condition \eqref{real22b} reduces to
\beq \label{11(11)_2}
\epsilon_{A}\epsilon_{B}=\bar{\eta}_{(A} \bar{\kappa}_{B)}.
\eeq
Since $\epsilon \neq 0$, this gives
\beqn
\epsilon_A= \alpha \, \bar{\eta}_A+ \beta \, \bar{\kappa}_A.
\eeqn
The arguments used for the analysis of type $1\underbar{111}$ solutions apply to give
\beqn
\epsilon \propto \bar{\eta}, \quad \text{or} \quad \epsilon \propto \bar{\kappa}.
\eeqn
As before, both these conditions give that $\eta \propto \kappa$, which contradicts the assumption that the solution is type $11\underbar{11}$.

\medskip

\noindent {\bf Type $2\underline{11}$}

\medskip

The Weyl spinor of type $2\underbar{11}$ solutions is of the form
\beq
C_{ABCD}=\epsilon_{(AB} \eta_{C} \eta_{D)},
\eeq
where $\epsilon_{AB}$ does not factorise.  Reality condition \eqref{real22b} gives that $\epsilon_{AB}$ factorises, contradicting the assumption that the solution is type $2\underbar{11}$.  Reality condition \eqref{real22a} gives a Majorana condition on $\eta$, since it implies that
\beqn
\eta_{A}\eta_{B}=\bar{\eta}_{A} \bar{\eta}_{B}.
\eeqn
Thus, type $2\underbar{11}$ solutions are also not possible.

\begin{figure}[ht]
\begin{center}
\begin{pspicture}(9.4,3.5)(0,0)
\rput(0,1){4} \rput(1.38,0){31} \rput(1.36,2){22}
\rput(2.91,1){211} \rput(5.05,1){1111} \rput(5.05,2){$\underbar{22}$}
\rput(7.18,1){$\underbar{11} \, \underbar{11}$}
\rput(9.32,1){$\scriptstyle{C_{abcd}\equiv 0}$}
\psline[linewidth=0.2mm]{->}(0.2,0.9)(1.1,0.05)
\psline[linewidth=0.2mm]{->}(0.2,1.1)(1.1,2)
\psline[linewidth=0.2mm]{->}(1.65,0.05)(2.55,0.9)
\psline[linewidth=0.2mm]{->}(1.65,2)(2.55,1.1)
\psline[linewidth=0.2mm]{->}(3.25,1)(4.55,1)
\psline[linewidth=0.2mm]{->}(5.5,1)(6.6,1)
\psline[linewidth=0.2mm]{->}(7.7,1)(8.6,1)
\psline[linewidth=0.2mm](1.36,2.28)(1.36,2.3)
\psline[linewidth=0.2mm]{->}(5.05,2.3)(5.05,2.28)
\psline[linewidth=0.2mm](1.76,2.7)(4.66,2.7)
\psline[linewidth=0.2mm]{->}(7.18,1.6)(7.18,1.28)
\psline[linewidth=0.2mm](5.35,2)(6.79,2)
\psarc[linewidth=0.2mm](1.75,2.3){0.4}{90}{180}
\psarc[linewidth=0.2mm](4.66,2.3){0.4}{0}{90}
\psarc[linewidth=0.2mm](6.79,1.6){0.4}{0}{90}
\end{pspicture}
\end{center}
\caption{Revised figure showing the 8 different algebraic types in the spinor classification.}
\label{fig:smet2}
\end{figure}
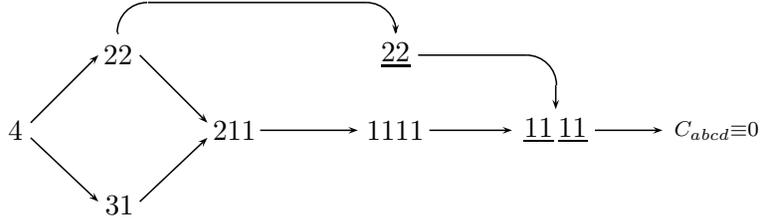

In summary, we have shown that, the reality condition on the Weyl spinor means that a solution cannot be of types $\underline{1111}$, $1\underbar{111}$, $11\underline{11}$ and $2\underbar{11}$.  Therefore, the number of possible types reduces to eight.  A revised version of figure \ref{fig:smet} is drawn in figure \ref{fig:smet2}.

Although, we considered the spinor classification of Lorentzian solutions in the analysis above, the spinor classification of Euclidean solutions is identical except that one uses the Euclidean Clifford algebra.  The reason for this is that as with the 5d Lorentzian Clifford algebra, the 5d Euclidean Clifford algebra does not admit a Majorana representation.  This means that the Weyl spinor of the Euclidean solution must satisfy the same reality condition as the Lorentzian case, i.e. equation \eqref{reality}, where $A$ now defines a Majorana condition for the Euclidean Clifford algebra.  However, since our arguments do not depend on specific properties of $A$, the same conclusions as those found above will follow.

\subsection{Relation to Petrov classification} \label{smet4d}

The five dimensional De Smet classification is a generalisation of the four dimensional Petrov classification, insofar as it is concerned with the factorisability of a totally symmetric 4-spinor that is equivalent to the Weyl tensor.  Here, we discuss the relation between the two classification schemes.

The 4d analogue of the De Smet Weyl spinor is
\beq \label{smetspi4}
C_{ABCD}= C_{abcd} {\gamma^{ab}}_{AB} {\gamma^{cd}}_{CD},
\eeq
where $\gamma^a$ form a representation of the 4d Clifford algebra. We need to show that the 4d De Smet Weyl spinor defined above is totally symmetric.  We can do this by using the results found in 5d.

In the definition of the 5d Weyl spinor (equation \eqref{weyspi5}) restrict the indices to take values $0, \dots, 3$ so that
\beq \label{smet4:1}
C_{ABCD}=  C_{abcd} (\gamma_0 \gamma_5 \gamma^{ab})_{AB} (\gamma_0 \gamma_5 \gamma^{cd})_{CD},
\eeq
where we have used the Clifford algebra representation defined in \eqref{rep}, for which $C=\gamma_0 \gamma_5$.  Lower case Latin indices range now from 0 to 3. Using the definition of $\gamma_5$ we find that
\beqn
\gamma_5 \gamma^{ab}=\frac{1}{2} {\varepsilon^{ab}}_{ef} \gamma^{ef},
\eeqn
where $\varepsilon_{abcd}$ is the Levi-Civita or permutation tensor.  Given the relation \cite{zakharov}
\beqn
C_{efgh}=\frac{1}{4}{\varepsilon^{ab}}_{ef} {\varepsilon^{cd}}_{gh}C_{abcd},
\eeqn
equation \eqref{smet4:1} reduces to
\beqn
C_{ABCD}= C_{abcd} (\gamma_0 \gamma^{ab})_{AB} (\gamma_0 \gamma^{cd})_{CD}.
\eeqn
Choosing the four-dimensional charge conjugation matrix, $C=\gamma_0$ gives the 4d De Smet Weyl spinor
\beq
C_{ABCD}= C_{abcd} {\gamma^{ab}}_{AB} {\gamma^{cd}}_{CD},
\eeq
which must be totally symmetric since the spinor we began with is totally symmetric.

In 4d, the Weyl tensor has 10 independent components, while a general totally symmetric 4-spinor has 35 complex independent components. However, the definition of the 4d De Smet Weyl spinor using the 4d Weyl tensor in equation \eqref{smetspi4} ensures that it has 10 complex independent components.  Put another way, the symmetries of the Weyl tensor in 5d give that the Weyl spinor is totally symmetric, whereas in 4d, the symmetries give more constraints on the spinor, including the condition that it be totally symmetric.  The fact that the Weyl tensor is real further constrains the De Smet Weyl spinor via a reality condition that halves its number of real independent components to 10.

In the Petrov classification, the homomorphism between $SL(2,\mathbb{C})$ and the Lorentz group is used to relate chiral spinor and Lorentz indices
\beq
X_{\alpha \dot{\alpha}}= i X^a \sigma_{a \, \alpha \dot{\alpha}}, \qquad X^a=\frac{i}{2}X_{\alpha \dot{\alpha}} \bar{\sigma}^{a \, \dot{\alpha} \alpha},
\eeq
where $\alpha$ and $\dot{\alpha}$ are left-handed and right-handed chiral spinor indices, respectively, $\sigma^a=(1,\sigma^i)$ and $\bar{\sigma}=(1,-\vec{\sigma})$.

Using the symmetries of the Weyl tensor and spinor calculus, it can be shown that the spinor equivalent of the Weyl tensor
\beq \label{petweyl}
C_{\alpha \beta \gamma \delta \dot{\alpha} \dot{\beta} \dot{\gamma} \dot{\delta}}={\sigma^a}_{\alpha \dot{\alpha}} {\sigma^b}_{\beta\dot{\beta}} {\sigma^c}_{\gamma \dot{\gamma}} {\sigma^d}_{\delta \dot{\delta}} C_{abcd},
\eeq
is equivalent to a totally symmetric spinor $\Psi_{\alpha \beta \gamma \delta}$, known as the \emph{Weyl spinor} in the Petrov classification \cite{pen,stew} 
\beq \label{weylspi4}
C_{\alpha \beta \gamma \delta \dot{\alpha} \dot{\beta} \dot{\gamma} \dot{\delta}}=\Psi_{\alpha \beta \gamma \delta} \varepsilon_{\dot{\alpha}\dot{\beta}} \varepsilon_{\dot{\gamma}\dot{\delta}}+\varepsilon_{\alpha \beta} \varepsilon_{\gamma \delta}\bar{\Psi}_{\dot{\alpha} \dot{\beta} \dot{\gamma} \dot{\delta}},
\eeq
where $\varepsilon_{\alpha \beta}=\varepsilon_{\dot{\alpha}\dot{\beta}}$ are the alternating tensors, which can be used to lower undotted and dotted indices, i.e. they act as charge conjugation matrices for chiral spinors.  Now, multiplying equation \eqref{petweyl} with $\varepsilon^{\dot{\alpha}\dot{\beta}}=-\varepsilon_{\dot{\alpha}\dot{\beta}}$, and contracting over dotted indices gives
\beq \label{petspi}
\Psi_{\alpha \beta \gamma \delta}=C_{abcd} {\zeta^{ab}}_{\alpha \beta} {\zeta^{cd}}_{\gamma \delta},
\eeq 
where equation \eqref{weylspi4} has been used and
\beq
{\zeta^{ab}}_{\alpha \beta}=\frac{1}{2} \varepsilon^{\dot{\alpha}\dot{\beta}} {\sigma^a}_{\alpha \dot{\alpha}} {\sigma^b}_{\beta \dot{\beta}}
\eeq
are the Lorentz algebra generators.  The fundamental theorem of algebra guarantees that $\Psi_{\alpha \beta \gamma \delta}$ factorises
\beq \label{petrov}
\Psi_{\alpha \beta \gamma \delta}=\alpha_{(\alpha} \beta_{\beta} \gamma_{\gamma} \delta_{\delta)}.
\eeq
The Petrov classification concerns the multiplicity of the factors in \eqref{petrov}, with the Petrov types defined in table \ref{tab:pet} \cite{steph}.

\begin{table}[ht]
\caption{The Petrov classification of the Weyl tensor} \centering
\vspace{1.5mm}
\begin{tabular}{c c c}
\hline\hline
Petrov type & Multiplicities & Diagram \\ [0.5ex]
\hline I &(1,1,1,1) &
\begin{pspicture}(-0.2,0.2)(0.8,0.9)
\psline[linewidth=0.3mm]{->}(0,0)(0,0.7)
\psline[linewidth=0.3mm]{->}(-0.015,0)(0.7,0)
\psline[linewidth=0.3mm]{->}(0,0)(0.60621,0.35)
\psline[linewidth=0.3mm]{->}(0,0)(0.35,0.60621)
\end{pspicture}\\
II & (2,1,1) &
\begin{pspicture}(-0.2,0.2)(0.8,0.9)
\psline[linewidth=0.3mm,doubleline=true,doublesep=0.15mm]{->}(0,0)(0,0.7)
\psline[linewidth=0.3mm]{->}(0.015,0)(0.60621,0.35)
\psline[linewidth=0.3mm]{->}(0.015,0)(0.35,0.60621)
\psline[linewidth=0.3mm,linecolor=white](0,-0.018)(0.5,-0.018)
\psline[linewidth=0.1mm,linecolor=white](0,0)(0,0.2)
\end{pspicture}\\
D & (2,2) &
\begin{pspicture}(-0.2,0.2)(0.8,0.9)
\psline[linewidth=0.3mm,doubleline=true,doublesep=0.15mm]{->}(0,0)(0,0.7)
\psline[linewidth=0.3mm,doubleline=true,doublesep=0.15mm]{->}(0.025,0)(0.49497474,0.49497474)
\psline[linewidth=0.3mm,linecolor=white](0,-0.016)(0.5,-0.016)
\psline[linewidth=0.1mm,linecolor=white](0,0)(0,0.2)
\end{pspicture}\\
III & (3,1) &
\begin{pspicture}(-0.2,0.2)(0.8,0.9)
\psline[linewidth=0.3mm,doubleline=true,doublesep=0.66mm]{->}(0,0)(0,0.7)
\psline[linewidth=0.3mm]{->}(0,0)(0,0.7)
\psline[linewidth=0.3mm]{->}(0.043,0)(0.49497474,0.49497474)
\psline[linewidth=0.3mm,linecolor=white](0,-0.016)(0.5,-0.016)
\end{pspicture}\\
N & (4) &
\begin{pspicture}(-0.2,0.2)(0.8,0.9)
\psline[linewidth=0.3mm,doubleline=true,doublesep=1mm]{->}(0,0)(0,0.7)
\psline[linewidth=0.3mm,doubleline=true,doublesep=0.15mm]{->}(0,0)(0,0.7)
\psline[linewidth=0.3mm]{->}(0,0.36)(0,0.6)
\end{pspicture}\\
O & $(C_{abcd} \equiv 0)$ &
\begin{pspicture}(-0.2,0.2)(0.8,0.9)
\psline[linewidth=0.1mm](-0.1,0.3)(0.1,0.3)
\end{pspicture}\\ [1ex]
\hline
\end{tabular}
\label{tab:pet}
\end{table}

Going back to the De Smet Weyl spinor in four dimensions,
\beqn
C_{ABCD}= C_{abcd} {\gamma^{ab}}_{AB} {\gamma^{cd}}_{CD},
\eeqn
we work in a chiral representation given by
\beq \label{weylrep}
\gamma^{a}=\begin{pmatrix} 0 & \sigma^{a} \\ -\bar{\sigma}^{a} & 0  \end{pmatrix}
\eeq
in the hope of relating the De Smet Weyl spinor to the Petrov Weyl spinor $\Psi_{\alpha \beta \gamma \delta}$ as defined in \eqref{petspi}.  The reality condition is
\beq \label{real4}
C_{ABCD}=C_{\dot{A}\dot{B}\dot{C}\dot{D}} {A^{\dot{A}}}_{A}{A^{\dot{B}}}_{B}{A^{\dot{C}}}_{C}{A^{\dot{D}}}_{D},
\eeq
where
\beq \label{A4lo}
A=\begin{pmatrix} 0 & -i\sigma^{2} \\ i\sigma^{2} & 0  \end{pmatrix}.
\eeq
In the chiral representation, ${\gamma^{ab}}_{AB}$ are block diagonal.  Thus, using the fact that $C_{ABCD}$ is totally symmetric, we deduce that 
\beq
C_{ABCD}=(C^{\alpha \beta \gamma \delta}, C_{\dot{\alpha} \dot{\beta} \dot{\gamma} \dot{\delta}}),
\eeq
where
\beqn
C_{\alpha \beta \gamma \delta}=C_{abcd} {\gamma^{ab}}_{\alpha \beta} {\gamma^{cd}}_{\gamma \delta},
\eeqn
i.e. mixed components vanish.  $C_{\alpha \beta \gamma \delta}$ and $C_{\dot{\alpha} \dot{\beta} \dot{\gamma} \dot{\delta}}$ are related via reality condition \eqref{real4}.  But, ${\gamma^{ab}}_{\alpha \beta}$=${\zeta^{ab}}_{\alpha \beta}$, and so 
\beq
C_{\alpha \beta \gamma \delta}=\Psi_{\alpha \beta \gamma \delta},
\eeq
i.e. the undotted part of the De Smet Weyl spinor in four dimensions is the Petrov Weyl spinor.  Equivalently, the Petrov polynomial
\beq
\Psi_{\alpha \beta \gamma \delta} \chi^{\alpha}\chi^{\beta}\chi^{\gamma}\chi^{\delta}=C_{ABCD} \psi^{A}\psi^{B}\psi^{C}\psi^{D},
\eeq
where $\psi=\begin{pmatrix} \chi \\ 0 \end{pmatrix}$.

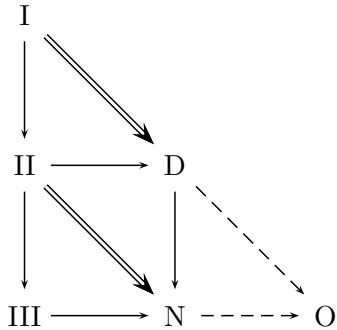
\begin{figure}[ht]
\begin{center}
\begin{pspicture}(4,4.5)(0,0.4)
\rput(0,0){III} \rput(2,0){N} \rput(4,0){O} \rput(0,2){II}
\rput(2,2){D} \rput(0,4){I}
\psline[linewidth=0.2mm]{->}(0,3.65)(0,2.35)
\psline[linewidth=0.2mm]{->}(0,1.65)(0,0.35)
\psline[linewidth=0.2mm]{->}(0.35,2)(1.65,2)
\psline[linewidth=0.2mm]{->}(0.35,0)(1.65,0)
\psline[linewidth=0.2mm]{->}(2,1.65)(2,0.35)
\psline[linewidth=0.2mm,linestyle=dashed]{->}(2.35,0)(3.65,0)
\psline[linewidth=0.2mm,doubleline=true,doublesep=0.4mm]{->}(0.28,3.72)(1.72,2.28)
\psline[linewidth=0.2mm,doubleline=true,doublesep=0.4mm]{->}(0.28,1.72)(1.72,0.28)
\psline[linewidth=0.2mm,linestyle=dashed]{->}(2.28,1.72)(3.72,0.28)
\end{pspicture}
\end{center}
\caption{Penrose diagram of the Petrov classification}
\label{fig:pen}
\end{figure}

We can move to a Majorana representation by performing a similarity transformation such that a Majorana spinor in the chiral representation
\beq \label{majspi}
\begin{pmatrix} \chi_{\alpha} \\ \bar{\chi}^{\dot{\alpha}} \end{pmatrix} \longrightarrow \sqrt{2} \begin{pmatrix} \text{Re}\chi \\ \text{Im}\chi \end{pmatrix}
\eeq
in the Majorana representation.  In a Majorana representation, the De Smet Weyl spinor is real.  Thus, the De Smet polynomial
\beq \label{desmet4d}
C(\psi)=C_{ABCD} \psi^A \psi^B \psi^C \psi^D,
\eeq
where $\psi$ is an arbitrary real (Majorana) spinor given by equation \eqref{majspi}, is real. Thus, the De Smet classification in 4d can be viewed as a classification of the Weyl tensor using Majorana spinors, in contrast to the Petrov classification, which uses chiral spinors.  Rewriting the polynomial as
\beqn
C(\psi)=C_{abcd} \gamma^{ab}(\chi) \gamma^{cd}(\chi),
\eeqn
where $\gamma^{ab}(\chi)={\gamma^{ab}}_{AB}\psi^A \psi^B$, it can be shown by direct calculation that
\beqn
\gamma^{ab}(\chi)=\zeta^{ab}(\chi)+\zeta^{ab}(\chi)^*,
\eeqn
where $\zeta^{ab}(\chi)={\zeta^{ab}}_{\alpha \beta} \chi^\alpha \chi^\beta$, so that
\beqn
C(\psi)=\Psi(\chi)+\Psi(\chi)^*+2C_{abcd} \zeta^{ab}(\chi)^* \zeta^{cd}(\chi),
\eeqn
where $\Psi(\chi)=C_{abcd}\zeta^{ab}(\chi)\zeta^{cd}(\chi)$ is the Petrov polynomial.  The tracefree property of the Weyl tensor implies
\beqn
C_{abcd} \zeta^{ab}(\chi)^* \zeta^{cd}(\chi)=0,
\eeqn
which implies
\beq \label{C4d}
C(\psi)=\Psi(\chi)+\Psi(\chi)^*.
\eeq

This can be used to relate De Smet types in four dimensions to Petrov types.  For example, assume that the solution is type N.  Then,
\beq
\Psi(\chi)=\omega^4.
\eeq
Equation \eqref{C4d} gives
\beq
C(\psi)=\omega^4+\omega^{* \, 4}=(\omega+\sqrt{-i} \, \omega^*)(\omega-\sqrt{-i} \, \omega^*)(\omega+\sqrt{i} \, \omega^*)(\omega-\sqrt{i} \, \omega^*).
\eeq
Hence, type N solutions are type 1111 in the 4d De Smet classification.  

We can consider the other Petrov types in a similar manner.  The results are summarised in table \ref{tab:petsmet}. What we find is that type I, II and III solutions are all algebraically general in the 4d De Smet classification, while type D and N solutions are type 22 and 1111, respectively.  Thus, the De Smet or Majorana spinor classification of the Weyl tensor in 4d is a coarse version of the Petrov or chiral spinor classification.

\begin{table}[ht]
\caption{Relation of De Smet classification in 4d to Petrov classification} \centering
\vspace{1.5mm}
\begin{tabular}{c c}
\hline
De Smet type & Petrov types \\ [0.5ex]
\hline 
4 & I, II, III \\
22 & D \\
1111 & N 
\\ \hline
\end{tabular}
\label{tab:petsmet}
\end{table}

The reason why other De Smet types are not possible goes back to the definition of the 4d De Smet Weyl spinor via the 4d Weyl tensor, in equation \eqref{smetspi4}. As discussed before, the symmetries of the 4d Weyl tensor imply not only that the spinor is totally symmetric, but give further conditions. It is these further conditions in addition to the reality condition that constrains the spinor in such a way that it can only admit three types.

For a Euclidean solution, we need to consider the Petrov classification of Euclidean solutions.  Starting from the chiral representation of the Lorentzian Clifford algebra \eqref{weylrep}, we can define a chiral representation for the Euclidean Clifford algebra by setting $\gamma^4=i\gamma^0$, i.e. $\gamma^{a}$ is given by
\beq
\gamma^{i}=\begin{pmatrix} 0 & \sigma^{i} \\ \sigma^{i} & 0  \end{pmatrix}, \quad \gamma^4=i\begin{pmatrix} 0 & 1 \\ -1 & 0  \end{pmatrix}
\eeq
where $i=$ 1, 2 or 3.  As with the Lorentzian case, since ${\gamma^{ab}}_{AB}$ are block-diagonal and $C_{ABCD}$ is totally symmetric, we conclude that
\beq
C_{ABCD}=(\Psi^{\alpha \beta \gamma \delta},\Psi_{\dot{\alpha} \dot{\beta} \dot{\gamma} \dot{\delta}}),
\eeq
where $\Psi_{\alpha \beta \gamma \delta}$ and $\Psi_{\dot{\alpha} \dot{\beta} \dot{\gamma} \dot{\delta}}$ are the spinor equivalents of the self-dual and anti-self-dual parts of the Weyl tensor.  Unlike the Lorentzian case, these two parts are independent of each other and this is manifested in the reality condition.  The reality condition on the Weyl spinor is
\beq
C_{ABCD}=C_{\dot{A}\dot{B}\dot{C}\dot{D}} {A^{\dot{A}}}_{A}{A^{\dot{B}}}_{B}{A^{\dot{C}}}_{C}{A^{\dot{D}}}_{D},
\eeq
where
\beq
A=\begin{pmatrix} -i\sigma^{2} & 0 \\ 0 & i\sigma^{2}  \end{pmatrix},
\eeq
i.e. for the Euclidean case, $A$, which defines the reality condition, is block-diagonal (compare this to \eqref{A4lo}).  Thus, rather than relating the two different parts of the Weyl spinor to one another, the reality condition places conditions on each part separately.  Analysing these conditions leads to the Petrov classification of Euclidean geometries \cite{karlhede}.

\begin{table}[ht]
\caption{De Smet classification and Petrov classification of 4d Euclidean metrics} \centering
\vspace{1.5mm}
\begin{tabular}{c c}
\hline
De Smet type & Petrov types \\ [0.5ex]
\hline 
4 & (I,I), (I,D), (D,I) \\
22 & (D,D) \\
1111 & (I,O), (O,I) \\
$\underline{11} \ \underline{11}$ & (D,O), (O,D)
\\ \hline
\end{tabular}
\label{tab:4deucl}
\end{table}

In the Petrov classification of Euclidean metrics, the two independent parts are classified separately and the Petrov type of the geometry is given as a pair consisting of the Petrov type of each part.  The reality condition implies that the Petrov types of the self-dual and anti-self-dual parts can only be I, D or O, leading to nine different cases.  The relation of these types to the De Smet types of the 4d geometry is given in table \ref{tab:4deucl}.

\subsection{Direct product solutions} \label{prod}

In \cite{typed},  direct and warped product solutions to vacuum Einstein equations are studied in the context of the CMPP classification and it is found that solutions with a one-dimensional Lorentzian factor are necessarily of types G, I, D or O. It is also found that solutions with a two-dimensional Lorentzian factor can only be of types D or O. 

Here, we study direct product solutions in the context of De Smet classification. The results below directly generalise to warped product manifolds for which the conformally related product manifold is again an Einstein manifold, since the Weyl tensors are conformally related. Let 
\beq
g_{ab}(x^a)=g_{\, \Gamma \Delta} (x^\Gamma) \oplus g_{\mu \nu}(x^\mu),
\eeq
where $g_{\, \Gamma \Delta}$ is the metric of a $n$-dimensional Lorentzian manifold, and $g_{\mu \nu}$ is the metric of a $(5-n)$-dimensional Euclidean manifold.  In such a setting, a tensor that splits like the metric---it has no mixed components and components belonging to one submanifold depend only on the coordinates covering that manifold---is known as a \emph{product-object} or as being \emph{decomposable}.  The Riemann tensor and its contractions are product-objects \cite{ficken}.  However, the Weyl tensor is in general not decomposable.

Assuming that the spacetime solves the vacuum Einstein equations
\beq \label{einstpro}
R_{ab}=\Lambda g_{ab} \qquad \iff \qquad R_{\, \Gamma \Delta}=\Lambda g_{\, \Gamma \Delta}, \ R_{\mu \nu}=\Lambda g_{\mu \nu},
\eeq
the mixed Weyl tensor components are
\begin{align}
C_{\Gamma \Delta \Theta \mu}=&C_{\Gamma \Delta \mu \nu}=C_{\Gamma \mu \nu \rho}=0, \notag \\
C_{\Gamma \mu \Delta \nu}&=-\frac{\Lambda}{4}g_{\, \Gamma \Delta} g_{\mu \nu}.
\end{align}
The non-mixed Weyl tensor components are
\begin{align}
 C_{\Gamma \Delta \Theta \Lambda}&=\begin{cases} \qquad \qquad \qquad \qquad 0, & n=1, \\ {}^{(1)}C_{\Gamma \Delta \Theta \Lambda} + \dfrac{(5-n)\Lambda}{2(n-1)} g_{\, \Gamma[\Theta}g_{\Lambda] \Delta}, & n \geq2, \end{cases} \notag \\
C_{\mu \nu \rho \sigma}&=\begin{cases}  {}^{(2)}C_{\mu \nu \rho \sigma} + \dfrac{n\Lambda}{2(4-n)} g_{\, \mu [\rho}g_{\sigma] \nu}, & \quad \; n \leq 4, \\\qquad \qquad \qquad \qquad 0, & \quad \; n=4, \end{cases}
\end{align}
where ${}^{(1)}C_{\Gamma \Delta \Theta \Lambda}$ and ${}^{(2)}C_{\mu \nu \rho \sigma}$ are the Weyl tensors derived from metrics $g_{\, \Gamma \Delta}$ and $g_{\mu \nu}$, respectively \cite{typed}.

For $n=1$ and $n=4$, the vacuum Einstein equations \eqref{einstpro} give that $\Lambda=0$.  For $n=4$, the Weyl spinor is
\beq
 C_{ABCD}={}^{(1)}C_{ABCD},
\eeq
i.e. the De Smet type of the solution is equal to the De Smet type of the four dimensional submanifold. Thus, from section \ref{smet4d}, we know that the De Smet type can only be one of 4, 22 or 1111. The relation between the De Smet type of the 5d solution and the Petrov type of the 4d factor is given in table \ref{tab:petsmet}.  The black string is an example of a direct product solution (with $n=4$). It is formed from the direct product of 4d type D Schwarzschild solution with a line, which means it is type 22 in De Smet classification \cite{desmet22}.

For $n=1$, the story is similar:
\beq
 C_{ABCD}={}^{(2)}C_{ABCD},
\eeq
and the De Smet type can only be one of 4, 22, 1111 or $\underline{11} \ \underline{11}$, as is shown in section \ref{smet4d}.  The relation between the De Smet type of the 5d solution and the Petrov type of the 4d factor is given in table \ref{tab:4deucl}.

For $n=2$, ${}^{(1)}C_{\Gamma \Delta \Theta \Lambda}={}^{(2)}C_{\mu \nu \rho \sigma}=0$.  Using a vielbein, the Weyl spinor is
\beq
C_{ABCD}=\frac{\Lambda}{4}\left[ 3 {\Gamma^{\Gamma \Delta}}_{AB}{\Gamma_{\Gamma \Delta}}_{CD}-4{\Gamma^{\Gamma \mu}}_{AB}{\Gamma_{\Gamma \mu}}_{CD}+2{\Gamma^{\mu \nu}}_{AB}{\Gamma_{\mu \nu}}_{CD} \right].
\eeq
If $\Lambda=0$, the solution is conformally flat.  If $\Lambda \neq 0$, the Weyl polynomial is 
\beq
C(\psi)=-24\Lambda (vw-uz)^2,
\eeq
where $\psi=(u,v,w,z)$.  Hence, such solutions are type $\underline{22}$.  This result mirrors that found in \cite{typed}, where it is shown that $n=2$ product solutions can only be types D and O.

For $n=3$, as with the $n=2$ case, we have ${}^{(1)}C_{\Gamma \Delta \Theta \Lambda}={}^{(2)}C_{\mu \nu \rho \sigma}=0$ and the Weyl spinor is
\beq
C_{ABCD}=\frac{\Lambda}{4}\left[  2{\Gamma^{\Gamma \Delta}}_{AB}{\Gamma_{\Gamma \Delta}}_{CD}-4{\Gamma^{\Gamma \mu}}_{AB}{\Gamma_{\Gamma \mu}}_{CD}+3{\Gamma^{\mu \nu}}_{AB}{\Gamma_{\mu \nu}}_{CD} \right].
\eeq
If the solution is not conformally flat, then $\Lambda \neq0$ and
\beq
C(\psi)=-24\Lambda (uw-vz)^2.
\eeq
Hence, such solutions are also type $\underline{22}$.

\section{Connection between tensor and spinor classifications}

\subsection{Connection between tensor and spinor classifications of two-form} \label{2formcomp}

A non-vanishing real two-form $F$ can be classified in two different ways. There is the tensorial approach analogous to the CMPP classification of the Weyl tensor whereby one looks for null vector $\ell$ such that in the null frame $(\ell,n,m^i)$ with $\ell \cdot n =-1$, $F_{0i}=0$. If this is the case, the 2-form is said to be aligned with respect to $\ell$ and of type I.  If no such $\ell$ exists, then the 2-form is of type G.  The 2-form is algebraically special (or type II) if $F_{0 i}=F_{01}=F_{i j}=0$.

We also have the spinorial approach outlined in section \ref{2form}, whereby the 2-form is algebraically special if and only if its bispinor equivalent $\epsilon_{AB}$ factorises, i.e. $\epsilon_{AB} = \epsilon_{(A} \bar{\epsilon}_{B)}$. The fact that the factors are conjugates of one another is implied by the reality condition, as shown in section \ref{2form}.  We call algebraically special 2-form fields type 11 fields in analogy with the De Smet classification of the Weyl tensor. Type 11 fields can be further classified by considering whether the scalar $f=\bar{\epsilon}\epsilon$ vanishes or not. If the bispinor does not factorise, we say that the field is type 2. 

The question we address in this section is how are the different types in the two classification schemes related to one another?

From Fierz identities \eqref{f1}--\eqref{f3}, we know that given a 2-form $F$ with bispinor equivalent $\e_{AB}$
\beqn
\epsilon_{AB}=\epsilon_{(A}\bar{\epsilon}_{B)} \qquad \implies \qquad V^a F_{ab}=0 \ \ \text{and} \ \ F^{ab}F_{ab}=-4 V^2
\eeqn
for $V^a=i  \bar{\e} \Gamma^a \e$.  For the case where $V$ is null, the converse can also be shown by direct computation, i.e. given a null vector $V$ and 2-form $F$ with bispinor equivalent $\e_{AB}$
\beqn
V^a F_{ab}=0 \ \ \text{and} \ \ F^{ab}F_{ab}=0 \qquad \implies \qquad \epsilon_{AB}=\epsilon_{(A}\bar{\epsilon}_{B)}.
\eeqn
We prove this by taking a general bispinor, finding its equivalent 2-form $F$ and showing that constraining it as on the left hand side above gives that the bispinor factorises. The reality condition gives that the factorisation must be of the form given on the right hand side and it turns out that $V^a \propto i  \bar{\e} \Gamma^a \e$.

From the above two results, we conclude that the 2-form field $F$ is
\begin{align}
\text{Type 11} \ (f=0) \quad \iff \quad \text{Type II}, \label{2form:2} \\ \label{2form:3}
\text{Type 11} \ (f \neq 0) \quad \implies \quad \text{Type G}.
\end{align}

Assume the field $F$ is type 2.  From \eqref{2form:2}, we know that the field cannot be type II.  However, there could still exist a null vector $\ell$ such that in the null frame $(\ell,n,m^i)$ $F_{0i}=0$, i.e. the field can be type I. If no such $\ell$ exists, then the 2-form is type G.

These results are summarised in table \ref{tab:2formtypes}.

\begin{table}[ht]
\caption{Relation between spinor and tensor types for 2-form} \centering
\vspace{1.5mm}
\begin{tabular}{c c}
\hline
Spinor type & Possible tensor types \\ [0.5ex]
\hline 
2 & G, I \\
11 & G, II
\\ \hline
\end{tabular}
\label{tab:2formtypes}
\end{table}

\subsection{Connection between CMPP and De Smet classifications} \label{CMPPSmet}

It is known that the definition of algebraic specialness in the CMPP and De Smet classification schemes do not agree \cite{her, axi}. Indeed, we showed in section \ref{prod} that the direct product of any type II or III 4d Ricci-flat solution with a line is algebraically general in the De Smet sense, but special in the CMPP sense.  Moreover, it is not even the case that the CMPP classification is a refinement, because there are examples that are algebraically special in the De Smet classification and general in the CMPP classification.  In this section, we investigate the connection between the two classification schemes.

\subsubsection{Relation of CMPP types to De Smet types} \label{CMPPtoSmet}

We shall proceed by assuming a five dimensional solution to be of particular CMPP type and consider what this means in the De Smet classification.  The vielbeins are chosen to be
\beqn
e_{\hat{0}}= \frac{1}{\sqrt{2}}(\ell+n), \quad e_{\hat{1}}= \frac{1}{\sqrt{2}}(\ell-n), \quad e_{\hat{i}}=m_i,
\eeqn
where $(\ell, n, m^i)$ form a null frame, such that $\ell$ and $n$ are null, $\ell \cdot n =-1$ and $m^i$ are a set of $d-2$ orthonormal spacelike vectors orthogonal to $\ell$ and $n$. The implicit Latin letters in the equation above label null frame indices. 

Now, using equations \eqref{weyspi5} and \eqref{Weylpol}, where $\psi=(u,v,w,z)$ we derive the Weyl polynomials associated with type D, III and N solutions and consider how they may factorise.

\medskip

\noindent {\bf Type N}

\medskip

For type N solutions the only non-zero components of the Weyl tensor are $C_{1i1j}$.  Using the trace-free property of the Weyl tensor ($C_{1i1}^{\ \ \ i}=0$), we have five independent Weyl tensor components: $C_{1313}, \ C_{1414}, \ C_{1213}, \ C_{1214},$ and $C_{1314}$.  Rotating $m_i \rightarrow m'_i$ such that $C_{1i1j}$ is diagonal in the new frame and computing the Weyl polynomial gives
\beq
C(\psi) = 8\left[ (2 C_{1313} + C_{1414}) (u^4+v^4) - 
 6 C_{1414} u^2 v^2 \right],
\eeq
which factorises to give
\beq \label{Nfac}
C(\psi) = A(u+a v)(u-a v)(u+v/a)(u-v/a),
\eeq
where $A=8(2 C_{1313} + C_{1414})$ and $a$ is given by
\beqn
a^2+1/a^2= \frac{6C_{1414}}{2 C_{1313} + C_{1414}},
\eeqn
assuming that $2 C_{1313} + C_{1414}\neq 0$. If $2 C_{1313} + C_{1414}=0$, then $C(\psi) \propto u^2 v^2$, which means the solution is type $\underline{11} \, \underline{11}$.  Equation \eqref{Nfac} implies that type N solutions are type 1111 or more special.  If $a=\pm \/1$, then the solution is of type $\underline{11} \, \underline{11}$. Hence, any type N solution must be of De Smet type 1111 or $\underline{11} \, \underline{11}$.

\medskip

\noindent {\bf Type III}

\medskip

For type III solutions, the non-zero components of the Weyl tensor are $C_{1i1j}$, $C_{1ijk}$ and $C_{011i}$.  The thirteen independent components are chosen to be: $C_{1212}, \ C_{1313},$ $C_{1213},$ $C_{1214}, \ C_{1314},$  $C_{1234}, \ C_{1342},$ $C_{1232}, \ C_{1242}, \ C_{1323}, \ C_{1343}, \ C_{1424}$ and  $C_{1434}$. Again, rotating the frame as before so that $C_{1i1j}$ is diagonal and computing the Weyl polynomial gives
\begin{align}
C(\psi) = & 8\left\{  (C_{1313}-C_{1212})(u^4 +v^4)+ 6( C_{1212}+ 
     C_{1313}) u^2 v^2 \right. \notag  \\  
&-\left. 
2\sqrt{2} w [(C_{1342}-C_{1234}+i C_{1343} - i C_{1242}) u^3 + 3(C_{1434}- i C_{1424})u^2 v   \right. \notag  \\  
&+\left.3(C_{1234}+ C_{1342}+iC_{1242}+iC_{1343})u v^2  +(2 C_{1232}+ C_{1434}+ 2 i C_{1323}+ i C_{1424}) v^3] \right. \notag \\ &+ \left. 
2\sqrt{2} z [(2C_{1232}+C_{1434}-2i C_{1323}-iC_{1424}) u^3-3(C_{1234}+ C_{1342}-iC_{1242}-iC_{1343})u^2 v \right. \notag  \\  
&+\left.3 (C_{1434}+i C_{1424} )u v^2  -(C_{1342}-C_{1234}+ i C_{1242}- i C_{1343}) v^3] \right\}.
\end{align}
Note that if all the coefficients of factors with $w$ or $z$ vanish, then this implies that the solution is type N giving a contradiction. It can be shown that the polynomial may factorise into cubic and linear factors. However, the conditions needed for the polynomial to factorise into two quadratic factors directly imply that one of the quadratic factors must factorise further into linear factors. The polynomial cannot be factorised any further without contradicting the assumption that the solution is type III.  Hence, any type III solution must be of De Smet type 4, 31 or 211.
\medskip

\noindent {\bf Type D}

\medskip

Finally, we consider type D solutions, for which the non-zero Weyl tensor components are $C_{ijkl}$, $C_{0i1j}$, $C_{01ij}$ and $C_{0101}$. The nine independent components are chosen to be: $C_{2323}, \ C_{2424},$ $C_{3434},$ $C_{0123}, \ C_{0124}, \ C_{0134}, \ C_{2324}, \ C_{3234}$ and  $C_{2434}$. We can rotate the spacelike basis vectors $m_i$ such that the symmetric part of $\Phi_{ij}\equiv C_{0i1j}$ is diagonal. Then, computing the Weyl polynomial gives
\begin{align}
C(\psi) = 16&\left\{  
 u^2 [3(\Phi_{22}-\Phi_{33}) w^2 +2(\Phi_{24}+ (2+i)\Phi_{34}) wz -(3(\Phi_{22}+\Phi_{33})+ 6i\Phi_{23}+4i\Phi_{24})z^2] \right. \notag  \\  
& + 2uv [-(\Phi_{24}+ (2+i)\Phi_{34})w^2 +6 \Phi_{44} wz + (\Phi_{24}+ (2-i)\Phi_{34})z^2] \notag  \\  
& -\left. v^2 [(3(\Phi_{22}+\Phi_{33})- 6i\Phi_{23}-4i\Phi_{24}) w^2 +2(\Phi_{24}+ (2-i)\Phi_{34}) wz -3(\Phi_{22}-\Phi_{33})z^2]  \right\}.
\end{align}
If the polynomial factorises, then, to avoid a contradiction, it does so into two non-factorisable quadratic factors that may or may not be proportional to one another, or into four independent linear factors. Hence, any type D solution must of De Smet type 4, 22, $\underline{22}$ or 1111.

\begin{table}[ht]
\caption{Possible De Smet types given CMPP type} \centering
\vspace{1.5mm}
\begin{tabular}{c c}
\hline
CMPP type & Possible De Smet types \\ [0.5ex]
\hline 
D & 4, 22, $\underline{22}$, 1111 \\
III & 4, 31, 211 \\
N & 1111, $\underline{11} \ \underline{11}$ 
\\ \hline
\end{tabular}
\label{tab:cmpptosmet}
\end{table}

\subsubsection{Relation of De Smet types to CMPP types} \label{sec:smettocmpp}

Now, we shall consider the different De Smet types and study what possible CMPP types they imply.  As with the previous section, we shall not be able to examine all De Smet types.  However, we shall study all algebraically special types, except type 31.  All such solutions can be regarded as special cases of type 22 solutions, for which the form of the Weyl tensor is given in equation \eqref{Weyl22}
\begin{align} 
C_{abcd}=&A_{a[c}B_{d]b}+B_{a[c}A_{d]b} -A_{ab}B_{cd}-B_{ab}A_{cd}-\frac{1}{2} A^{ef}B_{ef}g_{a[c}g_{d]b} \notag  \\ \label{Weyl22a} &-A_{ae}{B^e}_{[c}g_{d]b} -B_{ae}{A^e}_{[c}g_{d]b}+A_{be}{B^e}_{[c}g_{d]a}+B_{be}{A^e}_{[c}g_{d]a},
\end{align}
where 2-forms $A_{ab}=i\, tr(\Gamma_{ab} \epsilon)$ and $B_{ab}=i\, tr(\Gamma_{cd} \eta)$ satisfy one of the following reality conditions
\beq \label{rcon22a}
A^*_{ab}=A_{ab}, \quad B^*_{ab}=B_{ab}
\eeq
or
\beq \label{rcon22b}
A^*_{ab}=B_{ab}.
\eeq 

Also, we shall require results \eqref{2form:2} and \eqref{2form:3} and the generalisation of result \eqref{2form:2} for 2-forms that do not satisfy any reality condition.  If a 2-form $F$ is type I, then it can be shown by direct computation that
\beq \label{2form:1g}
\epsilon_{AB}=\epsilon_{(A}\kappa_{B)} \qquad \iff \qquad F \ \text{is type II},
\eeq
where $\epsilon_{AB}$ is the spinor equivalent of $F$.

\medskip

\noindent {\bf Type $22$}

\medskip

Working in null frame $(\ell, n ,m^i)$ such that $\ell \cdot n=-1$, the +2 boost weight components of the Weyl tensor $\Omega_{ij} \equiv C_{0i0j}$ are of the form
\beqn
\Omega_{ij}=A_{0k}B_{0k} \delta_{ij}-3A_{0(i}B_{|0|j)}.
\eeqn
$A_{0i}=0$ or $B_{0i}=0$ is sufficient for $\Omega_{ij}=0$.  However, one can show that it is also necessary. $\Omega_{ij}=0$ gives
\beq \label{omega0}
A_{0k}B_{0k} \delta_{ij}=3A_{0(i}B_{|0|j)}.
\eeq
Assume that neither $A_{0i}$ nor $B_{0i}$ vanish. Contracting the above equation with $A_{0i} A_{0j}$ gives
\beq \label{AB0}
A_{0k}B_{0k}=0.
\eeq 
Hence, from equation \eqref{omega0}
\beqn
A_{0(i}B_{|0|j)}=0.
\eeqn
Contracting this with $A_{0i}$ and using equation \eqref{AB0} gives
\beqn
A_{0i}=0.
\eeqn
But, this contradicts the original assumption that $A_{0i} \neq 0$. Therefore, either $A_{0i}=0$ or $B_{0i}=0$.

This means that the solution is type I or more special if and only if $A_{0i}=0$ or $B_{0i}=0$.  If there does not exist a $\ell$ such that $A_{0i}$ or $B_{0i}$ vanish then the solution is type G.  Now, assume there exists a $\ell$ such that without loss of generality $A_{0i}=0$.

Given that $A_{0i}=0$, the solution is type II if and only if $\Psi_{ijk} \equiv C_{0ijk}=0$.
\beqn
\Psi_{ijk}=B_{0[j}A_{k]i}-A_{jk}B_{0i}+A_{01} B_{0[j}\delta_{k]i}-B_{0l}A_{l[j}\delta_{k]i}.
\eeqn
It can be shown that
\beq
\Psi_{ijk}=0 \quad \iff \quad A_{01}=A_{ij}=0, \quad \text{or} \quad B_{0i}=0. 
\eeq
From the result in equation \eqref{2form:1g}, $A_{01}=A_{ij}=0$ would imply that bispinor $\epsilon$ factorises, contradicting the assumption that the solution is type 22.  Thus, $B_{0i}$ must also vanish for the solution to be type II.  This would be a further condition if reality conditions \eqref{rcon22a} are satisfied. However, $A_{0i}=0 \implies B_{0i}=0$ if reality condition \eqref{rcon22b} is satisfied.

Given that the solution is type II, then it can be shown that boost weight 0 components vanish if and only if
\beq
A_{01}=A_{ij}=0, \quad \text{or} \quad B_{01}=B_{ij}=0,
\eeq
which would imply (using \eqref{2form:1g}) that one of the bispinors associated with $A$ or $B$ factorises, contradicting the original assumption.

However, if there exists an $n$ such that $A_{1i}=B_{1i}=0$, then the results above apply directly to give that the solution is type D.  As before, $A_{1i}=0 \iff B_{1i}=0$ for reality condition \eqref{rcon22b}.

To summarise, type 22 solutions are of CMPP types G, I, II or D, depending on whether the 2-forms $A$ and $B$ are aligned.  An example of a type 22 solution that is algebraically general in the CMPP sense is the `homogeneous wrapped object' of \cite{smet} \cite{axi}.

\medskip

\noindent {\bf Type $\underline{22}$}

\medskip

A special case of type 22 is when $A \propto B$. In this case the solution is type $\underline{22}$ and the Weyl tensor is completely determined by 2-form $F$; using equation \eqref{Weyl22a}
\beq \label{Weyl(22)_}
C_{abcd}=2(F_{a[c}F_{d]b}- F_{ab}F_{cd}-F_{ae}{F^e}_{[c}g_{d]b}+F_{be}{F^e}_{[c}g_{d]a}) -\frac{1}{2} F^2g_{a[c}g_{d]b},
\eeq
where $F^2=F^{ef}F_{ef}$.  The reality condition is simply that $F$ is real.  Using the results derived when analysing type 22 solutions, the solution is type II if and only if there exists a $\ell$ such that $F_{0i}=0$.  If no such $\ell$ exists then the solution is type G.  If, in addition, there exists a $n$ such that $F_{1i}=0$, then and only then is the solution type D. Any further constraint on $F$ contradicts the original assumptions.

Thus, in summary, type $\underline{22}$ solutions can only be of CMPP type G, II and D.  An example of a type $\underline{22}$ solution is the 5d Myers-Perry solution \cite{smperry} (CMPP type D \cite{colperry}).  For this solution, the 2-form $F$ that squares to give the Riemann tensor is conformal to a test Maxwell field on it.

\medskip

\noindent {\bf Type $211$}

\medskip

The solution is type 211 if one of the bispinors, for example $\eta$, factorises.  Reality condition \eqref{rcon22b} contradicts the assumption that only one of the spinors factorises.  One of the reality conditions in \eqref{rcon22a} implies that $\eta_{AB}=\eta_{(A}\bar{\eta}_{B)}$ (using the results of section \ref{2form}), so that $B_{ab}=i \bar{\eta}\Gamma_{ab}\eta$.  We can also form a vector from $\eta$, $V^{a}=i\bar{\eta}\Gamma^{a}\eta$.  The result in \eqref{2form:2} gives that $B_{0i}=B_{01}=B_{ij}=0$ if and only if $V$ is null.  Thus, the analysis splits to two cases of $V$ timelike or null.

If $V$ is timelike, then the solution is type I if and only there exists a $\ell$ such that $A_{0i}=0$. Otherwise, result \eqref{2form:3} implies that the solution is type G.  Any other constraints on $A$ or $B$ contradict the original assumptions.

If $V$ is null, then choosing $\ell=V$ gives that the solution is type II.  If, in addition, $A_{0i}=0$ then solution is type III.  Any further constraints on the 2-forms give contradictions.

To summarise, type 211 solutions can only be of CMPP types G, I, II and III.

\medskip

\noindent {\bf Type $1111$}

\medskip 

For type 1111 solutions, both bispinors $\epsilon$ and $\eta$ factorise, i.e. $\epsilon_{AB}=\zeta_{(A}\kappa_{B)}$ and $\eta_{AB}=\lambda_{(A}\mu_{B)}$. Using arguments very similar to those used in section \ref{sec:smet} to show that type $11\underbar{11}$ solutions are not possible, one can show that reality condition \eqref{rcon22b} implies that at least two of the spinors coincide, contradicting the assumption that they are distinct.

Reality conditions \eqref{rcon22a} give that $\epsilon_{AB}=\epsilon_{(A}\bar{\epsilon}_{B)}$ and $\eta_{AB}=\eta_{(A}\bar{\eta}_{B)}$.  As above, we can form two vectors, $V^{a}=i\bar{\epsilon}\Gamma^{a}\epsilon$ and $W^{a}=i\bar{\eta}\Gamma^{a}\eta$.  

The solution is type G if and only if $V$ and $W$ are timelike. If only one of the vectors is null, then and only then is the solution type II.  The solution is type D if and only if both vectors are null but not proportional to one another. Finally, the solution is type N if and only if both vectors are null and proportional to one another.

Thus, type 1111 solutions are of CMPP types G, II, D or N.

\medskip

\noindent {\bf Type $\underline{11} \, \underline{11}$}

\medskip

Type $\underline{11} \, \underline{11}$ solutions are special cases of type 1111 solutions for which two pairs of spinors coincide.  However, it is more useful to think of them as special cases of type $\underline{22}$ solutions for which the bispinor $\epsilon$ factorises. The reality condition gives that $\epsilon_{AB}=\epsilon_{(A}\bar{\epsilon}_{B)}$.  Thus, the Weyl tensor is determined only from one spinor $\epsilon$.  Forming a vector from this, $V^{a}=i\bar{\epsilon}\Gamma^{a}\epsilon$, we find that the solution is type G if and only if $V$ is timelike and type N if and only if it is null.

The results found above are summarised in table \ref{tab:smettocmpp}.  These results are consistent with those found in section \ref{CMPPtoSmet} (see table \ref{tab:cmpptosmet}).

\begin{table}[t] 
\caption{Possible CMPP types given De Smet type} \centering
\vspace{1.5mm}
\begin{tabular}{c c}
\hline
De Smet type & Possible CMPP types \\ [0.5ex]
\hline 
22 & G, I, II, D \\
$\underline{22}$ & G, II, D \\
211 & G, I, II, III \\
1111 & G, II, D, N \\
$\underline{11} \ \underline{11}$ & G, N
\\ \hline
\end{tabular}
\label{tab:smettocmpp}
\end{table}

\subsection{De Smet classification of black ring} \label{br}

The results above show that a type G solution could be of any type in the De Smet classification.  This can be used to study algebraically general solutions as defined by the CMPP classification. The singly rotating black ring solution \cite{br} is a well-known example of a CMPP algebraically general five dimensional solution \cite{prav}.  Therefore, it would be desirable to know the De Smet type of the black ring solution.  The metric of the black ring can be written as \cite{empbrmet,bhigher}
\begin{align}
 ds^2=&-\frac{F(y)}{F(x)}\left( dt-C \, R \frac{1+y}{F(y)}d \psi\right)^2 \notag \\
  &+\frac{R^2}{(x-y)^2}F(x)\left(\frac{dx^2}{G(x)}-\frac{dy^2}{G(y)}+\frac{G(x)}{F(x)}d\phi^2-\frac{G(y)}{F(y)}d\psi^2 \right), \label{brmet}
\end{align}
where
\beqn
F(\zeta)=1+\lambda \zeta, \qquad G(\zeta)=(1-\zeta^2)(1+\nu \zeta), \qquad C=\sqrt{\lambda(\lambda-\nu)\frac{1+\lambda}{1-\lambda}},
\eeqn
The parameters $\lambda$ and $\nu$ are not independent and are related via an equation that will not be given here.  Furthermore, they satisfy $0<\nu \leq \lambda<1$. The coordinates $x$ and $y$ lie in the ranges $-1 \leq x \leq 1$ and $-\infty \leq y \leq -1$.  Asymptotic infinity is at $x=y= -1$, the ergosurface is at $y=-1/\lambda$ and inside this is the horizon at $y=-1/\nu$.

Choosing the following vielbein
\begin{align}
 e_{\hat{0}}=\sqrt{\frac{F(y)}{F(x)}}\left( dt-C \, R \frac{1+y}{F(y)}d \psi\right), \quad
e_{\hat{1}}=\frac{R}{x-y}\sqrt{\frac{F(x)}{G(x)}} dx, \quad e_{\hat{2}}=&\frac{R}{x-y}\sqrt{\frac{-F(x)}{G(y)}} dy, \notag \\
e_{\hat{3}}=\frac{R}{x-y}\sqrt{G(x)} d\phi, \quad 
e_{\hat{4}}=\frac{R}{x-y}\sqrt{\frac{-F(x)G(y)}{F(y)}}& d\psi, \label{brtet}
\end{align}
which is well-defined everywhere, except at the horizon and ergosurface, the Weyl polynomial is
\begin{align}
 C(\psi)=\frac{6(x-y)}{R^2 F(x)^3 F(y)}&\left\lbrace  A_1 (vw+uz)(uw-vz)+i(u^2-v^2+w^2-z^2)[A_2 (vw+uz) \right. \notag \\
&+A_3(uw-vz) ] +A_4(u^4+v^4+w^4+z^4)+A_5 uvwz \notag \\
&+A_6\left.(u^2 z^2+v^2w^2)+A_7(u^2w^2+v^2z^2)+A_8(u^2v^2+w^2z^2) \right\rbrace,
\end{align}
where $\psi=(u,v,w,z)$ and $A_i$ are expressions involving $x,\ y,\ \lambda$ and $\nu$ given in appendix \ref{brweyl}.

The polynomial above does not factorise, in general.  Thus, the solution is algebraically general in the De Smet sense. If we take the static limit $\nu \rightarrow \lambda$, for which $C=0$, the polynomial is again not factorisable, so the static black ring is also algebraically general.

\section{Classification of 2-forms and Weyl tensors defined by a spinor}

\subsection{Algebraically special 2-form fields} \label{em}

In four dimensions, the algebraic specialness of the Weyl tensor of a solution is related to the admittance by the solution of algebraically special electromagnetic fields. The Mariot-Robinson theorem \cite{mariot,*robinson, steph} states that in a null frame defined by null vector field $V$, a test null electromagnetic field $F$ has only negative boost weight components ($F$ is \emph{algebraically special}), if and only if $V$ defines a shear-free geodesic null congruence.  But the Goldberg-Sachs theorem \cite{gold} states that a vacuum solution admits a shear-free geodesic null congruence if and only if the solution is algebraically special.  Thus in 4d the algebraic specialness of a 2-form test field satisfying Maxwell equations and the algebraic specialness of the Weyl tensor coincide.  Considering the property of algebraically special $p$-form fields in higher dimensions could help clarify the status of a higher dimensional generalisation of the  Goldberg-Sachs theorem. Some progress has been made in this regard \cite{mrort, ghp} using the CMPP classification.

Here, we consider the algebraic specialness of a 2-form field in the context of the De Smet classification. We define a real 2-form field $F$ to be algebraically special if and only if its bispinor equivalent
\beq
\epsilon_{AB}=\frac{i}{8} F_{ab} {\Gamma^{ab}}_{AB},
\eeq
factorises, i.e.
\beq
\epsilon_{AB}=\eta_{(A}\epsilon_{B)}.
\eeq
From section \ref{2form}, we find that requiring $F$ to be real implies that $\eta_A \propto \bar{\epsilon}_A$ and hence
\beq
F_{ab}=i \bar{\epsilon} \Gamma_{ab} \epsilon.
\eeq
Furthermore, we can construct a scalar $f$ and vector $V$ (defined in \eqref{scavec}) that are related to eachother and $F$ via the Fierz identities (equations \eqref{f1}--\eqref{f4}).

The equations of motion for $F$ are given by
\begin{align}
 dF&=0, \\
d\star F&=\lambda F\wedge F,
\end{align}
where $\lambda=0$ corresponds to Maxwell theory and $\lambda=-2/\sqrt{3}$ corresponds to minimal supergravity \cite{minsugra}.  The analysis divides naturally into two cases of $f$ zero and non-zero, corresponding to $V$ null and timelike, respectively.

\subsection*{$V$ null}

For the null case, we showed in section \ref{2formcomp} that this is equivalent to the algebraic specialness of $F$ in the CMPP sense. Also, equation \eqref{f5} reduces to
\beqn
F \w F=0,
\eeqn
so that
\beq
d \star F=0,
\eeq
i.e. $F$ solves Maxwell equations.
From equation \eqref{f4}, we have
\beq
F=V \w W,
\eeq
for some 1-form $W$. Equations \eqref{f1} and \eqref{f6} imply that
\beqn
V\cdot W=0 \quad \text{and} \quad W^2=1,
\eeqn
respectively.

Such Maxwell fields have been studied in \cite{ghp}, where it is shown that $V$ is geodesic; $W$ is an eigenvector of the shear matrix and the wedge product of the rotation matrix with $W$ vanishes, i.e.
\beq
V \cdot \nabla V \propto V,
\eeq
and in null frame $(V,n,m^i)$, $\rho_{ij}=\nabla_j V_i$ satisfies 
\begin{align}
 \rho_{(ij)} W_{j}&=\frac{\rho}{2}W_{i}, \\
 \rho_{[ij]}&=\frac{1}{2} (Y \w W)_{ij},
\end{align}
for some 1-form $Y$.

\subsection*{$V$ timelike}

Taking Hodge dual and then exterior derivative of equation \eqref{f4} gives
\beq \label{dsf4}
\left[ \lambda F +d\left(\frac{V}{f}\right)\right]  \w F=0.
\eeq
The interior product of the above equation with $F$ implies
\beq \label{dv}
\left( 4 \lambda f^3+\iota_{F}dV\right) F=2f \left[ \iota_{V} d\left(\frac{V}{f}\right)\right] \w V,
\eeq
and the interior product of this equation with $V$ gives
\beq \label{Vgeo}
V \cdot \nabla V=\frac{V \cdot \nabla f}{f} \ V,
\eeq
i.e. $V$ defines a timelike geodesic congruence.

Taking exterior derivative and then Hodge dual of equation \eqref{f4} and using equations of motion gives
\beqn
df^a V_{[a}F_{bc]}=f\nabla^a(V_{[a}F_{bc]}),
\eeqn
which is equivalent to
\beqn
f \left( F_{c[a} \nabla^c V_{b]} +F_{c[a}\nabla_{b]} V^{c} \right) -df^c F_{c[a}V_{b]}-(V\cdot \nabla f) F_{ab}=0.
\eeqn
Interior product of this with $F$ gives
\beq \label{divV}
f \nabla \cdot V+V\cdot \nabla f=0.
\eeq

Introduce coordinates $(t, x^m)$ such that $V=\p/\p t$. The metric can then be written as
\beq
ds^2=-f^2\left(dt+\omega(x^p)\right)^2+f^{-1}h_{mn}(t,x^p)dx^m dx^n,
\eeq 
where the manifold with metric $h_{mn}$ will be referred to as the base space. $\omega$ is a 1-form with components only on the base space.

Equation \eqref{f3} implies that one can regard $F$ as a 2-form on the base space.  Then, equation \eqref{f6} gives
\beq \label{comstruc}
{F_m}^{p}{F_p}^{n}=-{\delta_m}^{n},
\eeq
where indices have been raised with respect to $h_{mn}$.  Thus, ${F_m}^{n}$ defines an almost complex structure with respect to $h_{mn}$. Moreover, the Bianchi identity on $F$ implies that $F_{mn}$ is closed, i.e.
\beq
^{(h)}\nabla_{[p}F_{mn]}=0.
\eeq
Thus, $h_{mn}$ is an almost-K\"ahler metric.

Defining
\beqn
\xi=f^{-1}V,
\eeqn
so that $\xi^2=-1$ and $\xi \cdot \nabla \xi =0$, the expansion, shear and rotation of the timelike congruence are
\begin{align}
 \theta&=\nabla_a \xi_b (g^{ab}+\xi^a \xi^b)=-2f^{-2} V \cdot \nabla f, \label{theta} \\
 \sigma_{ab}&=\nabla_{(a} \xi_{b)}-\frac{1}{4}\theta (g_{ab}+\xi_a \xi_b)=f^{-1}\nabla_{(a} V_{b)}-V_{(a}df_{b)}-\frac{1}{4}f^{-1}\theta h_{ab}=\frac{1}{2}f^{-2}(\mathcal{L}_{V}h_{ab}), \\
 \omega_{ab}&= -\nabla_{[a} \xi_{b]}=-f^{-1}\nabla_{[a} V_{b]},
\end{align}
respectively, where we have used equation \eqref{divV} in the first line above..

The Bianchi identity and equation \eqref{f3} give that
\beq
\mathcal{L}_{V}F=0.
\eeq
Rewriting equation \eqref{comstruc} as
\beq
F_{mn}F_{pq}h^{np}=h_{mq}
\eeq
and taking Lie derivative with respect to $V$ gives
\beqn 
{F_m}^{n}{F_q}^{p}(\mathcal{L}_{V}h_{np})=-(\mathcal{L}_{V}h_{mq})
\eeqn
or
\beq \label{shearcon}
{F_m}^{n}{F_q}^{p}\sigma_{np}+\sigma_{mq}=0,
\eeq
i.e. the $(1,1)-$part of the shear vanishes.

Therefore, for a timelike 2-form field, we have that the solution admits a timelike geodesic congruence and that the base space is almost-K\"{a}hler.  Furthermore, the $(1,1)-$part of the shear vanishes. 

\subsection{Classification of type $\underline{11} \, \underline{11}$ solutions} \label{1spin}

Type $\underline{11} \, \underline{11}$ solutions are significant in that their Weyl tensors are fully determined from a single Dirac spinor. In this section, we consider type $\underline{11} \, \underline{11}$ solutions and use the Bianchi identity to classify them.  The spinor $\epsilon$ that defines the type $\underline{11} \, \underline{11}$ solution can be used to form a real scalar, vector and 2-form as follows
\beq
f=\bar{\epsilon}\epsilon, \qquad V^a=i\bar{\epsilon} \Gamma^a \epsilon, \qquad F_{ab}=i\bar{\epsilon} \Gamma_{ab} \epsilon.
\eeq
As noted in section \ref{2form}, these quantities are related via Fierz identities \eqref{f1}--\eqref{f4}.

Simplifying the general form of the Weyl tensor of type 22 or more special solutions (equation \eqref{Weyl22}) using the Fierz identities, we find that the Weyl tensor of type $\underline{11} \, \underline{11}$ solutions is of the form 
\beq \label{weylt}
C_{abcd}=2\left(F_{a[c}F_{d]b}-F_{ab}F_{cd}+ V_a V_{[c}g_{d]b} -V_b V_{[c}g_{d]a}+f^2 g_{a[c}g_{d]b}\right).
\eeq
Assuming a vacuum Einstein solution
\beq
R_{ab}=\Lambda g_{ab},
\eeq
the Bianchi identity reduces to
\beq
C_{ab[cd;e]}=0.
\eeq
We shall use the above equation to find restrictions on the spinor.  As in section \ref{em}, the analysis divides into two cases of $V$ null and timelike.

\subsection*{$V$ null}

As shown in section \ref{sec:smettocmpp}, this class of type $\underline{11} \, \underline{11}$ solutions are type N in the CMPP classification ($V$ is a multiple WAND).  Type N solutions to vacuum Einstein equations have been studied before \cite{pravbian, ghp}.  In \cite{ghp}, it is shown that in the null frame $(V,n,m_i)$, the optical matrix $\rho_{ij}=\nabla_j V_i$ is of the form
\beqn
\rho_{ij}=\frac{1}{2}\begin{pmatrix} \rho & a & 0 \\ -a & \rho & 0 \\ 0&0&0 \end{pmatrix},
\eeqn
and that if the expansion $\rho=0$ then $a=0$, giving a Kundt solution.

The Fierz identities give
\beq
F=V \w W,
\eeq
for some $W$ satisfying $V \cdot W=0$ and $W^2=1$.  The Weyl tensor is of the form
\beq
C_{abcd}=-3F_{ab} F_{cd}+2(V_a V_{[c}g_{d]b} -V_b V_{[c}g_{d]a}).
\eeq
Then, $V^b {C^a}_{bcd;a}=0$ and $W^b {C^a}_{bcd;a}=0$ give
\beq
V \cdot \nabla V= (\nabla \cdot V) V, 
\eeq
i.e. $V$ is geodesic. This has been shown before \cite{pravbian, axi}.  Now, $F^{cd} C_{ab[cd;e]}=0$ gives
\beq
\nabla \cdot V=0,
\eeq
i.e. $\rho=0$, which means that we have a Kundt solution.

Choosing null frame $(V,n,W,m^{\hat{i}})$ where $\hat{i}=3,4$ and $V \cdot n=-1$, ${C^{a}}_{bcd;a}=0$ gives
\beq
2 L_{1 \hat{j}}-\tau_{\hat{j}}=0,
\eeq
where
\beq
\nabla_{a} V_{b}=L_{11} V_{a} V_{b}+L_{1i} \, m^{(i)}_a V_{b}+\tau_{i} \, V_{a} m^{(i)}_b.
\eeq
Then, $W^{a} C_{ab[cd;e]}=0$ gives
\beq \label{dW}
V \w dW=0,
\eeq
which implies
\beqn
V\cdot \nabla W \propto V, \qquad \text{and} \qquad W\cdot \nabla W \propto V,
\eeqn
where we have used $\rho_{ij}=0$.

The above equations give that
\beq
[V,W] \propto V,
\eeq
which, by Frobenius' theorem, gives that the distribution spanned by $\{V,W\}$ is integrable.  Moreover, using hyper-surface orthogonality of $V$ and equation \eqref{dW}, the dual formulation of Frobenius' theorem implies that the distribution spanned by $\{V,m^{\hat{i}}\}$ is also integrable.  We also have such a structure for the null case in section \ref{em}.

The Bianchi identity then reduces to
\beq
\nabla_a W_b=\frac{1}{3}(\tau_2-2L_{12}) \, {m^{(\hat{i})}}_a {m_{(\hat{i}) \, b}}-V_a \ n \cdot \nabla W_b-V_a \, V_b \, (n^c n^d \nabla_c W_d)-n^c \nabla_a W_c \ V_b.
\eeq

Furthermore, the only non-zero components of the Weyl tensor, $\Omega'_{ij}\equiv C_{1i1j}$ are of the form 
\beqn
\Omega'_{ij}=-3 W_i W_j+\delta_{ij},
\eeqn
i.e. $\Omega'_{ij}$ has two equal eigenvalues.

The information above does not seem to be strong enough to allow a complete classification of all such solutions.  In summary, type $\underline{11} \, \underline{11}$ solutions to vacuum Einstein equations with $f=0$ are type N Kundt solutions.  In addition, the WAND is in two orthogonal integrable null distributions of dimensions two and three and $\Omega'_{ij}$ has two equal eigenvalues.

\subsection*{$V$ timelike}

Contracting the Bianchi identity with the inverse metric and using the trace-free property of the Weyl tensor gives
\beqn
{C^{a}}_{bcd;a}=0.
\eeqn
Then $F^{bc} {C^{a}}_{bcd;a}=0$ gives
\beq \label{1spint:1}
3 f^2 {F^{a}}_{d;a}+(f df^a-V \cdot \nabla V^{a}) F_{ad}=0,
\eeq
and $V^{c} {C^{a}}_{bcd;a}=0$ gives
\begin{align}
f^2 \nabla_{a} V_{b}+&({F^c}_a{F^d}_b+2{F^c}_b {F^d}_a)\nabla_{c} V_{d} \notag \\ \label{1spint:2}
-&\nabla \cdot V(f^2 g_{ab}+V_a V_b)+f \, V \cdot \nabla f \, g_{ab}+V_a V \cdot \nabla V_b-fdf_a V_{b}=0.
\end{align}
Using the Fierz identities and equation \eqref{1spint:1} to simplify $F^{ab}F_{ea;b}$, $F^{ab} F^{cd} C_{ab[cd;e]}=0$ reduces to 
\beq \label{1spint:3}
3 \, V \cdot \nabla f+ f \, \nabla \cdot V=0,
\eeq
and a geodesity condition on $V$
\beq \label{1spint:4}
f \, V \cdot \nabla V=(V \cdot \nabla f) V \, .
\eeq
Using equation \eqref{1spint:2} to simplify $V^{a} F^{cd} C_{ab[cd;e]}=0$ gives
\beq
\nabla_{(a} (f^{-1}V)_{b)}=\frac{1}{4} \nabla \cdot (f^{-1}V) (g_{ab}+f^{-2}V_{a}V_{b})
\eeq
and $F^{ab}C_{ab[cd;e]}=0$ reduces to
\beq
d(f^{1/3}F)=\frac{2}{3} f^{-3} (V\cdot \nabla f)\, V\w (f^{1/3}F).
\eeq

To simplify the above equations, define new vector and 2-form
\beq
W=f^{-1} V  \quad \text{and} \quad J=f^{1/3}F.
\eeq
Then, the equations above give that $W$ forms a shear-free affinely parametrised geodesic congruence
\beq \label{Wgeoshear}
W^2=-1, \qquad W \cdot \nabla W=0, \qquad \nabla_{(a}W_{b)}=\frac{1}{4} \nabla \cdot W (g_{ab}+W_{a}W_{b}),
\eeq
while $J$ satisfies
\beq \label{Jmaxwell}
d\star J=0, \qquad dJ=\frac{2}{3} f^{-1} (W\cdot \nabla f)\, W\w J.
\eeq
Also, we have
\beq \label{1spint:dW}
dW_{ab}-f^{-8/3}{J^c}_a{J^d}_b dW_{cd}=0.
\eeq
Introduce coordinates $(t, x^m)$ such that $W=\p/\p t$. The metric can then be written as
\beq
ds^2=-\left(dt+\omega(x^p)\right)^2+f^{-4/3}k_{mn}(x^p,t)dx^m dx^n,
\eeq 
where the manifold with metric $k_{mn}$ is called the base space, and $\omega$ is a 1-form with components only on the base space.  $J$ lives on the base space and equation \eqref{f6} gives that ${J_m}^n$ is an almost complex structure on it, i.e.
\beq
{J^m}_p{J^p}_n=-{\delta^m}_n,
\eeq
where base space indices have been raised using $k^{mn}$, and will be done so hereafter.  Equation \eqref{1spint:dW} implies that $dW$ is a $(1,1)-$form.
Written in terms of base space components, $V^e C_{ab[cd;e]}=0$ reduces to
\beq \label{1spint:dW2}
k_{m[p}dW_{q]n}-k_{n[p}dW_{q]m}-dW_{mr} {J^r}_{[p}J_{q]n}+dW_{nr} {J^r}_{[p}J_{q]m}=0.
\eeq
Contracting the above equation with $k^{mp}$ gives
\beq \label{11part}
dW_{nq}+{J^m}_{n} {J^p}_{q}dW_{mp}+(dW_{mp} J^{mp})J_{nq}=0.
\eeq
Now, contracting this with $J^{nq}$ gives
\beqn
dW_{mp} J^{mp}=0.
\eeqn
Hence, from equation \eqref{11part}, we have
\beq
dW_{mn}+{J^p}_{m} {J^q}_{n}dW_{pq}=0,
\eeq
i.e. the $(1,1)-$part of $dW$ vanishes. But, we have from before that $dW$ is a $(1,1)-$form. This means that
\beq
dW=0.
\eeq
Remaining coordinate freedom can then be used to set
\beq
\omega(x^p)\equiv 0.
\eeq 

Finally, $F^{bc} C_{ab[cd;e]}=0$ gives
\beq
{}^{(h)}\nabla_{p}J_{mn}=0,
\eeq
i.e. 2-form $J$ is a K\"ahler form on the base space.  

Now, taking the exterior derivative of the second equation in \eqref{Jmaxwell} and using equations \eqref{Jmaxwell} and \eqref{f4} gives
\beqn
d(f^{-1} W\cdot \nabla f)^a J_{ab}=0
\eeqn
or
\beqn
d(f^{-1} W\cdot \nabla f) \propto W.
\eeqn
Note that equation \eqref{1spint:3} gives that the expansion of the congruence defined by $W$
\beq \label{expansionf}
\theta=-4f^{-1} W\cdot \nabla f.
\eeq
Thus,
\beqn
d\theta = -(W \cdot \nabla \theta) \, W,
\eeqn
i.e. the expansion depends only on $t$:
\beq \label{expdep}
\theta=\theta(t).
\eeq

Since $W^a C_{abcd}=0$, we can think of the Weyl tensor as living on the base space
\beq
C_{mnpq}=-2 f^{-2/3} \left(J_{mn}J_{pq}-J_{m[p}J_{q]n} - k_{m[p}k_{q]n}\right). 
\eeq
Equation \eqref{f5} gives that
\beq
\varepsilon_{mnpq}=\frac{1}{2} (J \w J)_{mnpq},
\eeq
where $\varepsilon_{mnpq}=f^{8/3}W^a\varepsilon_{amnpq}$ is the Levi-Civita tensor on the base space. Equation \eqref{f4} gives that $J$ is self-dual
\beq \label{jselfd}
\star_{4}J=J.
\eeq
Using the above two equations gives
\beq
\star_{4}C_{mnpq}=\frac{1}{2} {\varepsilon_{mn}}^{rs}C_{rspq}=C_{mnpq},
\eeq
i.e. the 5d Weyl tensor with components restricted to the base space with metric $k$ is self-dual.

Rewriting the metric as
\beqn
ds^2=-dt^2+\tilde{h}_{mn}(t,x^p)dx^mdx^n,
\eeqn
the final equation that gives the shear-free property of the congruence defined by $W$ in \eqref{Wgeoshear} reduces to
\beqn
\partial_t \tilde{h}_{mn}=\frac{1}{2} \theta(t) \, \tilde{h}_{mn},
\eeqn
where we have used the fact that $W$ is hyper-surface orthogonal and the result in \eqref{expdep} that the expansion is only time-dependent.  The equation above can be integrated for each component to give
\beq
\tilde{h}_{mn}(t,x^p)=A(t)^2 h_{mn}(x^p),
\eeq
for some function $A(t)$.

Thus, the metric can be written as
\beq \label{5dmet}
ds^2=-dt^2+A(t)^2 h_{mn}(x^p)dx^mdx^n,
\eeq
where the expansion
\beq
\theta(t)=4 \, A'(t)/A(t),
\eeq
and $h$ is a conformally K\"{a}hler metric.

Defining $L$ by 
\beq
\Lambda=\frac{4 \epsilon}{L^2}, \qquad \epsilon \in \{-1,0,1\},
\eeq
Raychaudhuri's equation
\beqn
W \cdot \nabla \theta=-\frac{1}{4} \theta^2+\Lambda,
\eeqn
reduces to
\beqn
A''(t)-\, \frac{\epsilon}{L^2} A(t)=0.
\eeqn
The first integral of this equation gives
\beq \label{Aeqn}
A'(t)^2-\frac{\epsilon}{L^2}A(t)^2=-\eta,
\eeq
where we can use the freedom in soaking up constants into metric $h$ to normalise $\eta \in \{-1,0,1\}$.

Using equation \eqref{Aeqn}, Einstein equations reduce to a constraint on the Ricci tensor of metric $h$
\beq \label{baseein}
{}^{(h)}R_{mn}=3 \eta \ h_{mn},
\eeq
i.e. $h$ is an Einstein metric.  

Solving equation \eqref{Aeqn} gives
\begin{center}
\begin{tabular}{l|lll}
$A(t)$ & $\eta=-1$ & $\eta=0$ & $\eta=1$ \\
\hline
$\epsilon=-1$ & $L \sin (t/L)$ \\
$\epsilon=0$ &  $t$ & 1 \\
$\epsilon=1$ &  $L \sinh (t/L)$ & $e^{\pm t/L} \ $ & $L \cosh (t/L)$
\end{tabular}
\end{center}

Furthermore, because of the warped product nature of the metric and the duality property of the 5d Weyl tensor derived above, we can learn something about the duality property of the Weyl tensor of metric $h$.  The Weyl tensor of the 5d metric \eqref{5dmet} is proportional to the Weyl tensor of the conformally related direct product metric
\beqn
ds^2=-d\tilde{t}^2+h_{mn}(x^p)dx^mdx^n,
\eeqn
where $d \tilde{t}=dt/A(t)$.  Using equations derived in \cite{typed} and equation \eqref{baseein} one can show that
\beq \label{weylh}
C_{mnpq}={}^{(h)}C_{mnpq}-\eta \, h_{m[p}h_{q]n}.
\eeq
Above, we showed that the 5d Weyl tensor with indices restricted to the 4d base space with metric $k$ is self-dual.  The Weyl tensor in the equation above is the 5d Weyl tensor with indices restricted to 4d base space with metric $h$.  By finding the relation between these two Weyl tensors, one can show that the self-duality result above translates to
\beqn
\star(C_{mnpq}+\eta \, h_{m[p}h_{q]n})=C_{mnpq}+\eta \, h_{m[p}h_{q]n},
\eeqn
or
\beq
\star {}^{(h)}C_{mnpq}={}^{(h)}C_{mnpq},
\eeq
i.e. the manifold with metric $h$ has self-dual Weyl tensor.  Since $h$ is conformal to a K\"ahler metric $k$ with K\"ahler form $J$ self-dual (equation \eqref{jselfd}), this means that the self-dual part of ${}^{(h)}C_{mnpq}$ is type D \cite{kahD}.

A simpler way of deriving this result is to absorb the Euclidean indices in equation \eqref{weylh} by multiplying with \textit{gamma}-matrices so that
\beq
C_{ABCD}={}^{(h)}C_{ABCD}-\eta \, {\Gamma_{mn}}_{(AB}{\Gamma^{mn}}_{CD)},
\eeq
where we have used the definition of the Weyl spinor given in equation \eqref{weyspi5}.  By direct computation, or using the Fierz identity, one can show that
\beqn
{\Gamma_{mn}}_{(AB}{\Gamma^{mn}}_{CD)}=0,
\eeqn
so that
\beq \label{Wspinh}
C_{ABCD}={}^{(h)}C_{ABCD}.
\eeq
Since the solution is type $\underline{11} \, \underline{11}$, we find from the correspondence between the 4d De Smet classification and the Euclidean Petrov classification (table \ref{tab:4deucl}) that the solution with metric $h$ is self-dual and of Petrov type (D,O).

In summary, all type $\underline{11} \, \underline{11}$ ($f \neq 0$) solutions to vacuum Einstein equations are warped product solutions of the form
\beq \label{fmetric}
ds^2=-dt^2+A(t)^2 h_{mn}(x^p)dx^mdx^n,
\eeq
where $A(t)$ is one of the functions in the table above depending on the curvature of the 5d metric and the curvature of the 4d Euclidean metric $h$. The 4d manifold with metric $h$ is a self-dual Einstein solution \footnote{In 4d, self-dual Einstein solutions are quaternion-K\"{a}hler \cite{qkah}. The holonomy group of a quaternion-K\"{a}hler solution is a subgroup of $Sp(1)^2 \cong SU(2)^2$ \cite{berger}.}.  In addition, $h$ is conformal to a K\"ahler metric $k$, thus the solution is of Petrov type (D,O).  If $\eta=0$, $h$ is a self-dual Ricci-flat metric. Thus, it is hyper-K\"ahler (see e.g. \cite{solitons}).

Note that the $\epsilon=\eta=0$ case for which $A(t)=1$ corresponds to a direct product solution and agrees with the result found in section \ref{prod}, where it was shown that for type $\underline{11} \, \underline{11}$ direct product solutions, the Euclidean base space has Petrov type (D,O) or (O,D) depending on the choice of orientation.

It is simple to show the converse, i.e. that the uplift of all type (D,O) Einstein solutions with metric \eqref{fmetric} are type $\underline{11} \, \underline{11}$.  Equation \eqref{Wspinh} gives that the De Smet type of the 5d solution coincides with the De Smet type of the 4d base space.  Table \ref{tab:4deucl} gives that the De Smet type of the 5d solution is $\underline{11} \, \underline{11}$.  Therefore, we have that a solution is 
\begin{align*}
\text{type} \ \underline{11} \, \underline{11} \ (f\neq 0) \quad \iff \quad &\text{a cosmological solution with spatial geometry a} \\  &\text{type (D,O) Einstein solution (metric \eqref{fmetric})}.
\end{align*}

The Fubini-Study metric on $\mathbb{C}\mathbb{P}^2$ is an example of a $\eta>0$ type (D,O) Einstein solution. A $\eta<0$ example is the Bergman metric on complex hyperbolic space $\mathbb{C}\mathbb{H}^2$. An important example of a type D hyper-K\"ahler metric is the Euclidean Taub-NUT solution.  The $\Lambda=0$ 5d solution with metric \eqref{fmetric}, where $h$ is the metric of Euclidean Taub-NUT is a magnetic monopole solution of Kaluza-Klein theory \cite{gpsoliton, *sorkin}.

\bigskip

\begin{center} {\bf Acknowledgements} \end{center}

\noindent
I would like to thank H. S. Reall for suggesting this project and reading through a draft manuscript.  His many comments and suggestions have been invaluable throughout.  I would also like to thank M. Dunajski, M. Durkee, D. Kubiznak, H. Godazgar and S. Hervik for discussions. I am especially grateful to G. W. Gibbons for helpful discussions and advice, particularly with regard to the spinor classification and its relation to the Petrov classification.  I am supported by EPSRC.

\newpage
\appendix

\section{5d Clifford algebra} \label{cliff}
In five dimensions, the Clifford algebra is
\begin{equation}
    \{\Gamma_{a}, \Gamma_{b}\}=2 g_{a b}.
    \label{clifford}
\end{equation}
The \textit{gamma}-matrices have spinor index structure $(\Gamma_{a})^A_{\
\, B}$. Given a Dirac spinor $\psi$, we can define its Majorana and Dirac conjugates as 
\beq
\psi^{\text{C}}=\psi^{t}C \quad (\psi_B=\psi^A C_{AB})
\eeq
and
\beq
\bar{\psi}=\psi^{\dagger} B \quad (\bar{\psi}_A = \psi^{\dot{A}} B_{\dot{A}A})
\eeq
respectively, where $C$ is the charge conjugation matrix, defined by
\beq \label{charge}
\Gamma_{a}^t=C \Gamma_{a} C^{-1},
\eeq
and $B$ is the Dirac conjugation matrix, defined by
\beq \label{diraccon}
\Gamma_{a}^{\dagger}=-B \Gamma_{a} B^{-1}.
\eeq
Note that $\psi^{\dot{A}}\equiv \psi^{* A}$. 

It follows from Schur's lemma that $B$ and $C$ are unique up to a phase factor.  Moreover, $B$ is Hermitian or anti-Hermitian, where we are free to choose which and $C$ is antisymmetric \cite{lee}.  Also, from the definition of Clifford algebra, we find that $\Gamma_0$ is anti-Hermitian, while $\Gamma_i$ are Hermitian.  Thus, from equation \eqref{diraccon}, we find that $B$ is $\Gamma_0$ up to a phase.  We choose $B$ Hermitian so that assignment of indices is consistent, i.e. $\overline{B_{ \dot{A} A}}=(B^t)_{ \dot{A} A}=B_{A \dot{A}}$. 

The Majorana condition is
\beq
\bar{\psi}C^{-1}=\psi \quad \text{or} \quad \psi^{\ast}=A \psi  \quad (\psi^{\dot{A}}={A^{\dot{A}}}_{B} \psi^B),
\eeq
up to a phase, where $A=(C B^{-1})^t$.  In 5d, the Majorana condition has no non-trivial solutions, i.e. it implies that $\psi=0$.  This is because $A^* A=-1$.

A convenient representation to use for the five dimensional Clifford algebra is to start with the Majorana representation for the four dimensional Clifford algebra
\beq
\gamma_0=\begin{pmatrix} -i \sigma^2 & 0 \\ 0 & i \sigma^2 \end{pmatrix}, \
\gamma_1=\begin{pmatrix} \sigma^1 & 0 \\ 0 & \sigma^1 \end{pmatrix}, \
\gamma_2=\begin{pmatrix} \sigma^3 & 0 \\ 0 & \sigma^3 \end{pmatrix}, \
\gamma_3=\begin{pmatrix} 0 & i \sigma^2  \\ -i \sigma^2 & 0 \end{pmatrix},
\eeq
where $\sigma^{\hat{i}}$ for $\hat{i}=1,2,3$ are the usual Pauli matrices and add $\gamma_5= \gamma_0 \gamma_1 \gamma_2 \gamma_3$.  Thus,
\beq \label{rep}
\Gamma_a=(\gamma_0,\gamma_{\hat{i}},i \gamma_5).
\eeq 
Then
\beq
B=i \gamma_0, \qquad C= \gamma_0 \gamma_5 \qquad \text{and} \qquad A=-i \gamma_5.
\eeq

The five dimensional Fierz identity is
\beq \label{fierz}
M_{AB}N_{CD}=\dfrac{1}{4}C_{AD} (NM)_{CB}+\dfrac{1}{4} {\Gamma_{e}}_{AD}(N\Gamma^{e}M)_{CB}-\dfrac{1}{8} {\Gamma_{ef}}_{AD}(N\Gamma^{ef}M)_{CB}.
\eeq

For brevity, we omit factors of $C$ and $C^{-1}$ where it is clear that indices have been lowered or raised.

\section{Solving reality condition for 2-form} \label{reality2form}
In this appendix, we outline how one can prove that the reality condition 
\beq \label{real2form}
\e_{(A}\t_{B)}=\bar{\e}_{(A}\bar{\t}_{B)}
\eeq
implies
\beq
\t =\bar{\e},
\eeq
where $\e$ and $\t$ are non-zero spinors.  Letting $B=1,2,3,4$ in \eqref{real2form} gives four equations of the form
\beqn
\e_1 \t_A+\t_1 \e_A=\bar{\e}_1 \bar{\t}_A+\bar{\t}_1 \bar{\e}_A.
\eeqn

If all four of the equations are dependent, that is they are proportional to one another, then by considering all possible cases one can show that \footnote{In fact it is enough for only a pair of the equations above to be proportional to one another for this result to hold. We shall not use this, since we would like the sketch of the proof in this section to mirror that given in appendix \ref{reality22}.}
\beqn
\e \propto \t.
\eeqn
Then, \eqref{real2form} becomes
\beqn
\e_{(A}\e_{B)}\propto \bar{\e}_{(A}\bar{\e}_{B)}.
\eeqn
Since $\e$ is non-zero, assume without loss of generality that $\e_1 \neq 0$.  Then letting $B=1$ in the equation above gives
\beqn
\bar{\e} \propto \e,
\eeqn
which is a Majorana condition on $\e$.  The Majorana condition has no non-zero solutions in five dimensions.  Therefore, $\e \propto \t$ implies a contradiction.

Now, assuming that two of the equation are independent gives
\beq \label{bart}
\bar{\t}= \alpha \ \e +\beta \ \t,
\eeq
where $\alpha$ and $\beta$ are constants and $\alpha \neq 0$, otherwise we have a Majorana condition on $\t$.  Taking complex conjugate of the above equation and multiplying appropriately by $A$ gives
\beq \label{bare}
\bar{\e}=-\frac{1}{\alpha^*}\left( \alpha \beta^* \ \e+(1+|\beta|^2) \ \t \right).
\eeq

Substituting equations \eqref{bart} and \eqref{bare} into equation \eqref{real2form} gives
\beqn
\alpha^2 \beta^* \ \e_{(A}\e_{B)}+(\alpha +\alpha^*+2\alpha |\beta|^2) \ \e_{(A}\t_{B)}+\beta (1+|\beta|^2) \t_{(A}\t_{B)}=0.
\eeqn
Letting $B=1,2,3,4$ gives four equations relating $\e$ and $\t$.  Now, the analysis splits into three cases. The first case is that some coefficients in the equations are non-zero. This implies $\e \propto \t$, which gives a contradiction as shown above. The second case is that all coefficients in each of the equations vanish.  But, it can be shown that this too implies that $\e \propto \t$.  Thus, we are left with the final case that the coefficients in the equation above vanish.  Since $\alpha \neq 0$, we have $\beta=0$.

Then, equation \eqref{bart} gives $\bar{\t} \propto \e$ or
\beq
\t \propto \bar{\e}.
\eeq

\section{Solving reality condition for type 22} \label{reality22}
In this appendix, we outline how one can prove that the reality condition for type 22 solutions
\beq \label{real22}
\e_{(AB}\t_{CD)}=\bar{\e}_{(AB}\bar{\t}_{CD)},
\eeq
implies either
\beq
\e_{AB} = \bar{\e}_{AB}, \qquad \t_{AB} = \bar{\t}_{AB},
\eeq
or
\beq
\e_{AB} = \bar{\t}_{AB},
\eeq
where $\e$ and $\t$ are non-zero bispinors, and
\beqn
\bar{\e}_{AB}\equiv \e_{\dot{A} \dot{B}} {A^{\dot{A}}}_{A} {A^{\dot{B}}}_{B}.
\eeqn

The strategy used for the proof here is similar in nature to that used to prove the result in appendix \ref{reality2form}, except that there are more cases to consider.

First consider the case $\t \propto \e$.  Then, equation \eqref{real22} becomes
\beq \label{realeta}
\e_{(AB}\e_{CD)}=\bar{\e}_{(AB}\bar{\e}_{CD)}.
\eeq
Assume $\e_{11}\neq 0$.  This is true unless $\e_{AA}=0$ for all $A$ (no sum on $A$), in which case, it can be shown that $\e_{AB} \propto \bar{\e}_{AB}$, as required. Letting $A=B=C=D=1$ in equation \eqref{realeta} gives
\beqn
{\e_{11}}^2={{\bar{\e}_{11}}}^{\ \ 2}.
\eeqn
Now, letting $B=C=D=1$ in equation \eqref{realeta} and using the equation above gives
\beqn
\e_{A1}=\pm \bar{\e}_{A1}.
\eeqn
Finally, letting $C=D=1$ in equation \eqref{realeta} and using the two equations above gives
\beq
\e_{AB} = \bar{\e}_{AB}.
\eeq
Thus, we have proved the result above for $\t \propto \e$.  Note that this is equivalent to solving the reality condition for type $\underline{22}$ solutions.

Now, assume $\t \not\propto \e$.  Letting $B, \ C, \ D=1,2,3,4$ in equation \eqref{real22} with at least two of them coinciding gives 16 equations of the form
\beqn
\e_{A1}\t_{11}+\e_{11}\t_{A1}=\bar{\e}_{A1}\bar{\t}_{11}+\bar{\e}_{11}\bar{\t}_{A1},
\eeqn
or
\beqn
\e_{A2} \t_{11}+\e_{A1}\t_{21}+\e_{21}\t_{A1}+\e_{11} \t_{A2}=\bar{\e}_{A2} \bar{\t}_{11}+\bar{\e}_{A1}\bar{\t}_{21}+\bar{\e}_{21}\bar{\t}_{A1}+\bar{\e}_{11} \bar{\t}_{A2}.
\eeqn
If three of the 16 equations are not independent, then by considering all possible cases it can be shown that $\t \propto \e$, which contradicts the original assumption that $\t \not\propto \e$.  Therefore, at least 15 of the equations are independent.  They can be used to express $\e_{A2}, \ldots, \t_{A1}, \ldots, \bar{\e}_{A1}, \ldots, \bar{\t}_{A1}, \ldots, \bar{\t}_{A4}$ in terms of $\e_{A1}$.

Now, let $C=D=1,2,3,4$ in equation \eqref{real22} to give 4 equations of the form
\beqn
\e_{AB} \t_{11}+\e_{A1}\t_{B1}+\e_{B1}\t_{A1}+\e_{11} \t_{AB}=\bar{\e}_{AB} \bar{\t}_{11}+\bar{\e}_{A1}\bar{\t}_{B1}+\bar{\e}_{B1}\bar{\t}_{A1}+\bar{\e}_{11} \bar{\t}_{AB}.
\eeqn
Similar to before, if two of these four equations are not independent then it can be shown by considering all the different possibilities that $\t \propto \e$, which contradicts the original assumption.  Thus, three of the equations are independent, which means we can eliminate $\e_{A1}$, $\bar{\t}_{AB}$ and use the last equation to show that
\beq \label{bareps}
\bar{\e}= \alpha \ \e+ \beta \ \eta,
\eeq
where $\alpha$ and $\beta$ are constants and if $\beta=0$, it can be shown that
\beq
\bar{\t} \propto \t.
\eeq
Thus $\beta=0$ gives one of the possibilities allowed above: $\bar{\e} \propto \e$ and $\bar{\t} \propto \t$.  Furthermore, $\alpha=0$ gives the second possibility: $\bar{\e} \propto \t$.  Using equations \eqref{bareps} and \eqref{real22}, one can show that
\beq \label{alpha0}
\alpha=0 \iff |\beta|^2= 1 \iff \beta=\pm 1,
\eeq
and
\beq \label{beta0}
\beta=0 \iff |\alpha|^2= 1 \iff \alpha=\pm 1.
\eeq

Assume $\alpha \neq 0$ and $\beta \neq 0$.  Taking complex conjugate of equation \eqref{bareps} and multiplying appropriately by a pair of $A$'s gives
\beq \label{bareta}
\bar{\t}=\frac{1}{\beta^*}\left( (1-|\alpha|^2) \ \e - \alpha^* \beta \ \t \right).
\eeq
Substituting equations \eqref{bareps} and \eqref{bareta} into equation \eqref{real22} gives
\beq \label{real22et}
\lambda \ \e_{(AB}\e_{CD)}+\mu \ \e_{(AB} \t_{CD)}+\nu \ \t_{(AB}\t_{CD)}=0,
\eeq
where
\beqn
\lambda= \alpha (1-|\alpha|^2), \qquad \mu=\beta-\beta^*-2\beta |\alpha|^2, \qquad \nu=-\alpha^* \beta^2.
\eeqn
Using the first equivalence in \eqref{beta0} gives that $\lambda=0$ implies that $\alpha=0$ or $\beta=0$, which contradicts the original assumption.  This is also trivially true for $\nu=0$. $\mu=0$ implies
\beqn
\beta(1-2|\alpha|^2)=\beta^*.
\eeqn
Multiplying the equation above with its complex conjugate gives
\beqn
|\alpha|^2 |\beta|^2=0.
\eeqn
Thus, $\mu=0$ also contradicts the original assumption that $\alpha \beta \neq 0$.

Letting $B=C=D=1,2,3,4$ in equation \eqref{real22et} gives four equations of the form
\beqn
(2\lambda \e_{11}+\mu \t_{11})\ \e_{A1} +(2nu \t_{11}+\mu \e_{11})\ \t_{A1}=0.
\eeqn
The equations are independent unless the coefficients vanish.  If this is the case, then we have
\beqn
\mu^2-4\lambda \nu=0.
\eeqn
It can be shown that this implies that $\alpha =0$, which contradicts the original assumption.  Thus, by considering different components of the four equations, one can show that, in general, they imply that $\t \propto \e$, which contradicts the original assumption.  Although, one must also consider special cases, where, for example, $\e_{A1} \neq 0$ only for $A=1$. However, in these cases too, one can show that $\t \propto \e$.

Therefore, $\alpha \beta \neq 0$ contradicts the original assumption, which implies that $\alpha \beta =0$.

\section{Weyl tensor of type 22 or more special solutions} \label{weyl22s}
In this appendix, we use the 5d Fierz identity to derive the form of the Weyl tensor of type 22 or more special solutions.  For all such solutions, the Weyl spinor is of the form
\beq \label{psi22}
C_{ABCD}=\epsilon_{(AB}\eta_{CD)}.
\eeq
We can invert the definition of the Weyl spinor (equation \eqref{weyspi5}), so that given a Weyl spinor, the associated Weyl tensor is given by
\beq \label{weytenspi}
C_{abcd}=\frac{1}{64}(\Gamma_{ab})^{AB}(\Gamma_{cd})^{CD}C_{ABCD}.
\eeq

Using the form of the Weyl spinor \eqref{psi22} and equation \eqref{weytenspi}, the Weyl tensor is
\beq \label{weyl22}
C_{abcd}=-\frac{1}{2}(A_{ab} B_{cd}+B_{ab} A_{cd})+2 tr(\Gamma_{ab} \epsilon \Gamma_{cd} \eta),
\eeq
where 
\beqn
A_{ab}=i\, tr(\Gamma_{ab} \epsilon) \quad \text{and} \quad B_{ab}=i\, tr(\Gamma_{cd} \eta),
\eeqn
and $\epsilon$ and $\eta$ have been rescaled.

Using the 5d Fierz identity (equation \eqref{fierz}) with $M=\Gamma_{ab}$, $N=\Gamma_{cd}$ and using the fact that $C$ and $C\Gamma^a$ are antisymmetric in their spinor indices, while $C\Gamma^{ab}$ and $\eta$ are symmetric gives
\beq \label{fierz:22}
tr(\Gamma_{ab} \epsilon \Gamma_{cd} \eta)=\frac{i}{8} B_{ef} tr(\Gamma_{cd} \Gamma^{ef} \Gamma_{ab} \epsilon).
\eeq
Using the Fierz identity with $M=\Gamma_{cd}$, $N=\Gamma^{ef}$, contracting two spinor indices between $\Gamma_{cd}$ and $\Gamma^{ef}$, and multiplying by $\Gamma_{ab}$ gives an expression for $(\Gamma_{cd} \Gamma^{ef} \Gamma_{ab})_{CB}$, which when inserted into the equation above gives
\begin{align*}
tr(\Gamma_{ab} \epsilon \Gamma_{cd} \eta)=\frac{i}{32} B_{ef} \left(-i \, A_{ab} tr(\Gamma^{ef} \Gamma_{cd})+tr(\Gamma_{g} \Gamma_{ab} \epsilon) \right. & tr(\Gamma^{ef} \Gamma^g \Gamma_{cd}) \notag \\ &\left. -\frac{1}{2} tr(\Gamma_{gh} \Gamma_{ab} \epsilon) tr(\Gamma^{ef} \Gamma^{gh} \Gamma_{cd})\right). 
\end{align*}
Again, using the Fierz identities in a similar way to that used to derive equation \eqref{fierz:22} and properties of \emph{gamma}-matrices in 5d, in particular that
\beq
tr(\Gamma^a \Gamma^b \Gamma^c \Gamma^d \Gamma^e)=-4i\varepsilon^{abcde},
\eeq
gives
\begin{align*}
tr(\Gamma_{ab} \epsilon \Gamma_{cd} \eta)=&-\frac{1}{4}\left( A_{ab}B_{cd}+B_{ab}A_{cd}+A^{ef}B_{ef}g_{a[c}g_{d]b} \right) +\frac{1}{2}(A_{a[c}B_{d]b}+B_{a[c}A_{d]b})\notag \\ &-\frac{1}{2}\left(A_{ae}{B^e}_{[c}g_{d]b} +B_{ae}{A^e}_{[c}g_{d]b}-A_{be}{B^e}_{[c}g_{d]a}-B_{be}{A^e}_{[c}g_{d]a}\right).
\end{align*}
Equation \eqref{weyl22} then gives
\begin{align} 
C_{abcd}=&A_{a[c}B_{d]b}+B_{a[c}A_{d]b} -A_{ab}B_{cd}-B_{ab}A_{cd}-\frac{1}{2} A^{ef}B_{ef}g_{a[c}g_{d]b} \notag  \\ \label{Weyl22:AB} &-A_{ae}{B^e}_{[c}g_{d]b} -B_{ae}{A^e}_{[c}g_{d]b}+A_{be}{B^e}_{[c}g_{d]a}+B_{be}{A^e}_{[c}g_{d]a},
\end{align}
i.e. the Weyl tensor of type 22 solutions is determined by two 2-forms $A$ and $B$. 

\section{Weyl polynomial of black ring} \label{brweyl}
The Weyl polynomial of the singly rotating black ring \eqref{brmet}, using the tetrad given in \eqref{brtet}, is
\begin{align}
 C(\psi)=\frac{6(x-y)}{R^2 F(x)^3 F(y)}&\left\lbrace  A_1 (vw+uz)(uw-vz)+i(u^2-v^2+w^2-z^2)[A_2 (vw+uz) \right. \notag \\
&+A_3(uw-vz) ] +A_4(u^4+v^4+w^4+z^4)+A_5 uvwz \notag \\
&+A_6\left.(u^2 z^2+v^2w^2)+A_7(u^2w^2+v^2z^2)+A_8(u^2v^2+w^2z^2) \right\rbrace,
\end{align}
where $\psi=(u,v,w,z)$ and
\begin{align}
A_1&=8(x-y)(1-\lambda)CF(y) \sqrt{G(x)} , \notag \\
A_2&=-4(x-y)(1-\lambda)CF(x) \sqrt{-G(y)}, \notag \\
A_3&=4F(x)F(y)\sqrt{-G(x)G(y)}, \notag \\
A_4&=(1+x \lambda)^2[(x-y)^2\nu+(y^2-1)\lambda-(1-x^2)y\lambda \nu-2(x-y)y^2\lambda \nu], \notag \\
A_5&=8\left\lbrace (1-\lambda^2)[2xy(\lambda-\nu)+(x^2+y^2)\nu]-\lambda(y^2-1)(1+x^2 \lambda^2)\right. \notag \\ &+2\lambda(1+y\lambda)(y\lambda-x^2)+\left.\lambda \nu x(1-x^2)(1+y\lambda)^2\right\rbrace , \notag \\
A_6&=2\left\lbrace 3 (x - y)^2 \nu + 2 x \lambda^2(1 - y^2)  + x^3 \lambda \nu (1+x\nu) + 2 x y \lambda^2 \nu -2 (x-y)^2 \lambda^2 \nu + \lambda(1+y \nu) \right. \notag \\
&-  3 x^2  y \lambda \nu (1+y \lambda) -2(x-y)^2 \lambda (1 - \lambda^2) + x^3 \lambda \nu - y^2 \lambda+  x^3 y  \lambda^3 \nu (x-y)+x^2 \lambda^3(1- y^2) \notag \\
&+ \left.  x^2 y  \lambda^3 \nu (1-xy)+  v^2 (1 + x \lambda)^2 [(x - y)^2 \nu + \lambda(1-y^2)+ y \lambda \nu(1-xy) + x y \lambda \nu(x-y)] \right\rbrace, \notag \\
A_7&=2\left\lbrace(x - y)^2 \nu + -4 y \lambda^2 (1 - x^2) +  2 x \lambda^2 (y^2 - 1) - 4 x y \lambda^2 \nu (1 - x^2)+  x^2 \lambda^2 \nu (x^2 - y^2)  \right. \notag \\ 
&-  2 x y \lambda^2 \nu (1 + x y) + 2 x y^3 \lambda^2 \nu  -  3 \lambda + y^2 \lambda -2 x \lambda \nu +  4 x^3 \lambda \nu - y \lambda \nu +  2 y^3 \lambda \nu (1 + x\lambda) \notag \\
&\left.  + 2 x^2 \lambda -  3 x^2 y \lambda \nu - 2 y^2 \lambda^3 +  x^2 \lambda^3 ( y^2 - 1) + 2 x^2 \lambda^3 -  x^2 y \lambda^3 \nu (1 - x^2) -  2 x y^2 \lambda^3 \nu (1 - x  y) \right\rbrace, \notag \\ \notag
A_8&=2 (1 + x \lambda)^2\left\lbrace (x - y)^2 \nu - \lambda (y^2 - 1) + 
   y\lambda\nu (1 - xy) + x y\lambda\nu (x - y)\right\rbrace ].
\end{align}

\bibliography{weyl}
\bibliographystyle{utphys.bst}

\end{document}